\documentclass[traditabstract]{aa} % for the abstract without structuration 
\usepackage{natbib}
\usepackage{graphicx}
\usepackage{pict2e}
\usepackage{txfonts}
\usepackage{subfigure}
 \newcommand{\mic}{$\mu$m}
\def\kms    {\ifmmode{{\rm \ts km\ts s}^{-1}}\else{\ts km\ts s$^{-1}$}\fi}
\def\msol   {\ifmmode{{\rm M}_{\odot}}\else{M$_{\odot}$}\fi}

\begin{document}
   \title{Molecular gas in the inner 0.7\,kpc-radius ring of M31}

   \subtitle{}

   \author{A.-L. Melchior
          \inst{1,2}
	 \and
          F. Combes\inst{1} 
          }
   \offprints{A.L.Melchior@obspm.fr}

   \institute{LERMA,
Observatoire de Paris, LERMA, UMR8112, 61, avenue de l'Observatoire, Paris, F-75
014, France\\
              \email{A.L.Melchior@obspm.fr,Francoise.Combes@obspm.fr}
\and
Universit\'e Pierre et Marie Curie-Paris 6, 4, Place Jussieu,
F-75\,252 Paris Cedex 05, France\\ }

   \date{}

% \abstract{}{}{}{}{} 
% 5 {} token are mandatory
 
\abstract {
The study of the gas kinematic in the central 1.5\,kpc$\times$1.5\,kpc
region of M31 has revealed several surprises. The starting point of
this investigation was the detection at the IRAM-30m telescope of
molecular gas with very large line splittings up to 260\,km\,s$^{-1}$
within the beam ($\sim$40\,pc). In this region, which is known for its
low gas content, we also detect an ionised gas outflow in the
circumnuclear region (within 75\,pc from the centre) extending to the
whole area in X-ray. Relying on atomic, ionised, and molecular gas, we
account for most observables with a scenario that assumes that a few
{hundreds} Myr ago, M\,31 underwent a frontal collision with M\,32,
which triggered some star-formation activity in the centre, and this
collision explains the special configuration of M\,31 with two rings
observed at 0.7\,kpc and 10\,kpc.  The inner disc (whose rotation is
detected in HI and ionised gas ([NII])) has thus been tilted
(inclination: 43$\deg$, PA: 70$\deg$) with respect to the main disc
(inclination: 77$\deg$, PA: 35$\deg$). One of the CO velocity
components is compatible with this inner disc, while the second one
comes from a tilted ring-like material with 40$\deg$ inclination and
PA=-35$\deg$. The relic star formation estimated { by previous works}
to have occurred more than 100\,Myr ago could have been triggered by
the collision { and could be linked to} the outflow detected in the
ionised gas.  Last, we demonstrate that the amplitude of the line
splittings detected in CO centred on the systemic velocity with a
relatively high spatial resolution (40\,pc) cannot be accounted for by
a possible weak bar that is roughly aligned along the minor
axis. Although M31 has a triaxial bulge, there are no bar indicators
in the gas component (photometry, no strong skewness of the
isovelocities, etc.).  }

\maketitle
%
%________________________________________________________________

\section{Introduction}
The merging processes that probably took place during the assembly of
the galaxies of the Local Group is now actively
studied \citep[e.g.][]{Klimentowski:2010}. While the Milky Way
has not suffered any major merger for several billion years and
presents only a series of dwarf galaxy stellar streams
\citep[see][]{Helmi:2008}, the other main spiral of the Local Group
(M31) has had a more perturbed history. Giant stellar loops and tidal
streams are observed in the surroundings, extending up to the
neighbour M33, which is suspected to have interacted with M31
\citep[e.g.][]{McConnachie:2009}. Many coherent structures observed
around M31 are interpreted as the disruption of small dwarf galaxies,
which are numerous in the vicinity of M31 \citep[e.g.][]{Ibata:2004}.

At the same time, it has been known for a long time that the M31 galaxy
exhibits an unusual morphology  
\citep[e.g.][]{Arp:1964,Haas:1998,Helfer:2003}.  The young stellar
population tracers in the disc are all concentrated in the so-called
10\,kpc ring { \citep{Chemin:2009, Nieten:2006,Azimlu:2011}}, which is
highly contrasted and superposed to only a few spiral structures {
\citep[e.g.][]{Nieten:2006,Gordon:2006}}. One striking feature is the
small amount of gas present in the central region: neither
\citet{Braun:2009} nor
\citet{Chemin:2009}  detect any significant HI component, while a
similar depletion is observed in CO-emission by
\citet{Nieten:2006}. It is usually described as a quiescent galaxy
with little star formation { SFR$\sim$0.4\,\msol\,yr$^{-1}$
\citep[e.g.][]{Barmby:2006,Tabatabaei:2010,Azimlu:2011} and} with an
ultra-weak nuclear activity \citep{DelBurgo:2000}. However, ionised
gas is detected in the central field
\citep[e.g.][]{Rubin:1971,Ciardullo:1988,Boulesteix:1987,Bogdan:2008,Liu:2010},
usually interpreted in term of shocks, and \citet{Melchior:2000} has
detected only a small amount of molecular gas {($1.5\times 10^4
\msol$)} within { 1.3\,\arcmin} (305\,pc in projection). 
{ There is no obvious on-going star formation in this central region
\citep[e.g.][]{Olsen:2006,Li:2009,Azimlu:2011} and the ionised gas 
\citep[approximately 1500\,\msol\, according to][]{Jacoby:1985} can be
accounted for by mass lost from evolving stars.}

In this ``empty'' central region, interest has focused on the
supermassive black hole \citep{Bacon:1994}, { its activity
\citep[e.g.][]{Li:2011}} and the circumnuclear region (inner few hundred
parsecs) { \citep{Li:2009}}. \citet{Barmby:2006} observed the whole
galaxy in the mid-infrared with the {\em Infrared Array Camera} {
(IRAC)} on board the {\em Spitzer Space Telescope}, revealing
spectacular dust rings and spiral arms. \citet{Block:2006} stress the
presence of an {elongated} inner ring with projected diameters
1.5\,kpc by 1\,kpc and propose a completely new interpretation for the
morphology of this galaxy. Both rings at 1\,kpc and 10\,kpc are
off-centred. Their respective radii do not correspond to what is
expected from resonant rings in a barred spiral galaxy. The most
likely scenario for the formation of these rings is a head-on
collision of the Cartwheel type {
\citep{Struck-Marcell:1993,Horellou:2001}}.  Unlike the Cartwheel,
where the companion is about 1/3rd of the mass of the target (major
merger), in Andromeda, the collision can be called a minor merger, and
produces much less contrasted rings in the main disc.
\citet{Block:2006} propose that the collision partner was M\,32, with
about 1/10th of the mass (dark matter included) at the
beginning. After stripping experienced in the collision, the M\,32
mass is now 1/23 that of the main target M\,31.  The M\,32 plunging
head-on through the centre of M\,31 has triggered the propagation of
an annular wave, which is now identified with the 10\,kpc {ring}, and
a second wave propagates more slowly behind
\citep[see e.g.][]{Appleton:1996}, and would correspond to the inner
ring.  In addition, the inner ring has formed {and is propagating} in
a tilted and warped disc, which accounts for its almost face-on
appearance, in contrast with the inclined main disc of M\,31.  This
scenario explains why the cold gas has been expelled from the central
region and also why there are shocks and hot gas.  In this article we
focus on the gas content of the inner ring and inside.

We present CO(1-0) and CO(2-1) observations obtained with the IRAM-30m
telescope in different positions located in the north-western part of
the inner ring, and a few positions inside the ring. While the CO
intensities are correlated in first instance with the A$_B$
extinction, the velocity distributions are quite unexpected. In the
inner ring the velocities are spread between -450\,\kms and -150\,\kms
for a few positions while the systemic velocity {(-310\,\kms)} is
expected in this area along the minor axis. To better understand these
velocities, we compared our measurements with the velocities available
for the ionised {and HI} gas {and} found that they do not really
match.  While neither the main HI warp of the disc {nor a bar along
the line of sight} can explain the wide amplitude velocity splittings
observed along the minor axis, the {main} configuration {compatible
with the data discussed here} is to have a ring inclined with respect
to the nuclear disc that would correspond in projection to the inner
dust ring, which is detected in extinction and in infra-red
emission. This ring could be attributed to a head-on collision with
M32 \citep{Block:2006} or to some accretion of gas from an M31-M33 gas
bridge or tidal loop detected in HI \citep{Braun:2004}.

In Section \ref{sect:obs} we describe the observations performed at
the IRAM-30m telescope. In Section \ref{sect:datared} we describe the
reduction of the C0 data. In Section \ref{sect:ana} we present an
analysis of the CO data and compare the molecular gas with the ionised
gas, together with information from other wavelengths. In Section
\ref{sect:inter} we explore various scenarios and propose one
modelling that explains the observations. In Section \ref{sect:disc} we
discuss our results and conclude.

%__________________________________________________________________

\section{Observations}
\label{sect:obs}
\begin{figure*}
\setlength{\unitlength}{1cm}
\begin{picture}(16,11) (0,0)
     \put(-0.1,5){
\begin{minipage}[l]{.2\linewidth}
\includegraphics[width=6.5cm,angle=0]{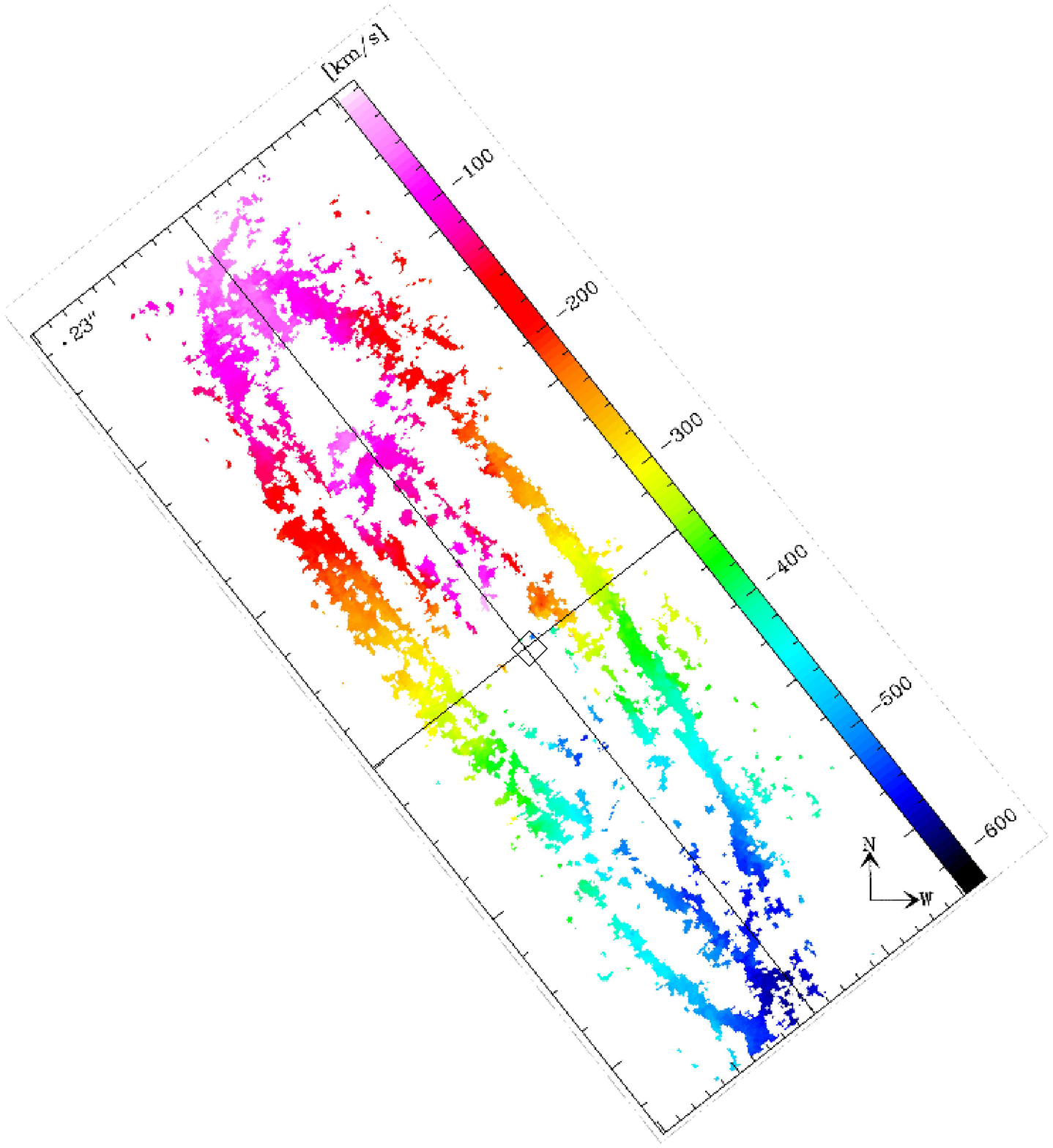} 
\end{minipage} \hspace{2.5cm}
\begin{minipage}[r]{.7\linewidth}
\vspace{0.1cm}
\includegraphics[height=11.5cm,angle=-90]{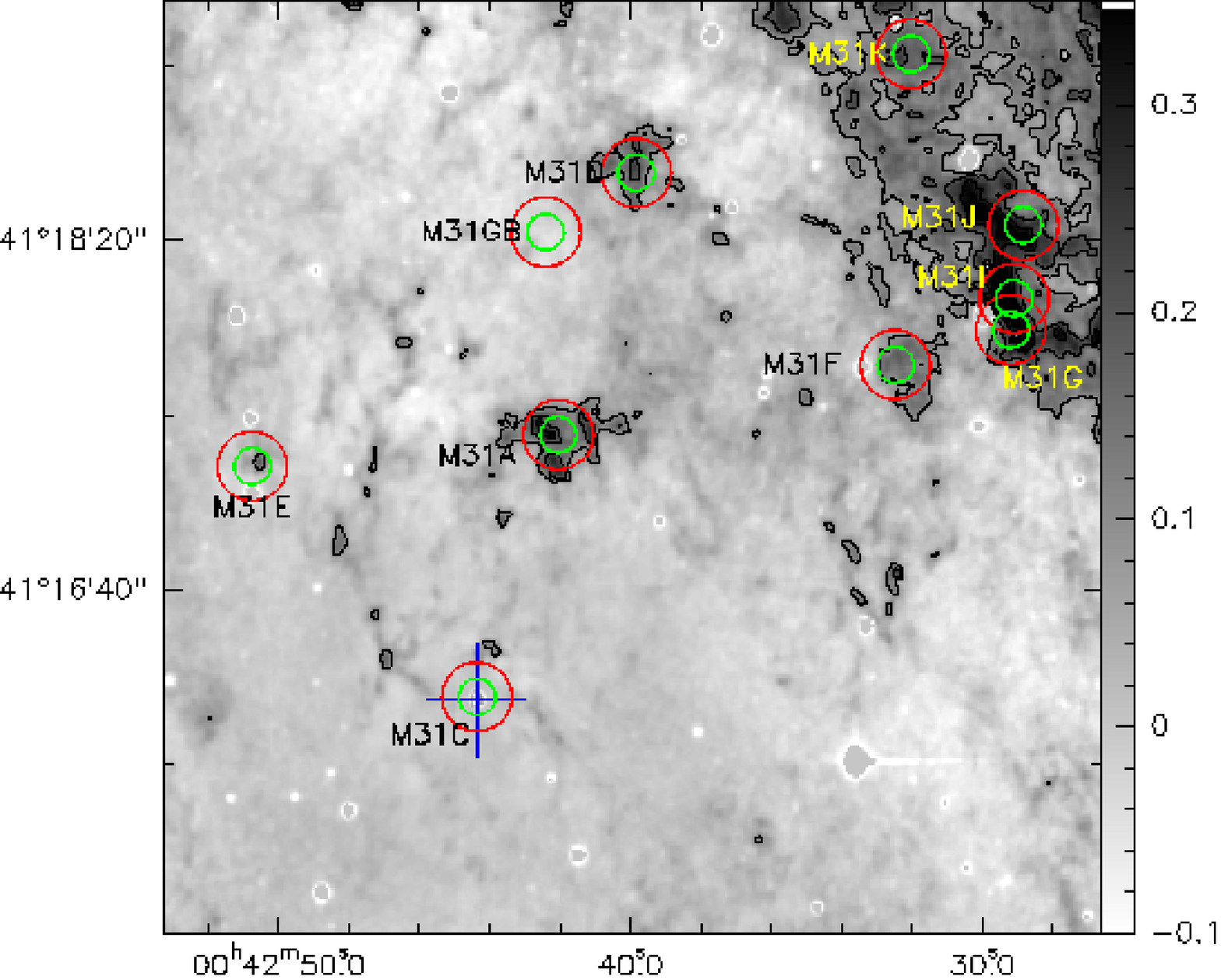}
\end{minipage}
}
\put(3.15, 4.65){\line(5, 6){4.18}}
\put(7.33, 9.65){\line(5, 0){9.77}}
\put(3.25, 4.50){\line(3, -3){3.73}}
\put(6.99, 0.77){\line(5, 0){10.12}}
%\put(6.99, 0.77){\line(5, 0){11.12}}
\end{picture}
\caption{{Right:} Positions {(J2000)} observed in CO in the
central part of M\,31. The circles indicate 1-mm (green) and 3-mm
(red) beams of the 10 positions observed superimposed on the $A_B$
extinction map obtained in \cite{Melchior:2000}. As discussed in this
paper, the extinction with the central 2 arcsec is not well-defined on
this map. The (blue) cross indicates the centre of M31
($\alpha_{J2000}$=$00^{h} 42^{m} 44^{s}.371$,
$\delta_{J2000}$=$41^{\circ} 16^{\prime} 08^{\prime\prime}.34$,
\cite{Crane:1992}.) {Left: Velocity field measured in CO for the
whole galaxy by \citet{Nieten:2006}. A small rectangle in the central
part indicates the field studied in this paper. This global view
recalls the large-scale configuration of this galaxy: a steeply
inclined disc at 77${\deg}$ with a position angle of 35${\deg}$.}}
\label{fig:super} 
\end{figure*}

%\begin{figure*}
% \centering 
%\begin{minipage}[c]{.2\linewidth}
%\includegraphics[width=6.5cm,angle=0]{rotatedok2.eps} 
%\end{minipage} \hfill
%\begin{minipage}[c]{.7\linewidth}
%\includegraphics[height=12.5cm,angle=-90]{super2.eps}
%\end{minipage}
%\caption{Positions observed in CO in the central part of M\,31. The
%circles indicate 1-mm (green) and 3-mm (red) beams of the 10 positions
%observed superimposed on the $A_B$ extinction map obtained in
%\cite{Melchior:2000}. As discussed in this paper, the extinction with
%the central 2 arcsec is not well defined on this map. The (blue) cross
%indicates the centre of M31 ($\alpha_{J2000}$=$00^{h} 42^{m}
%44^{s}.371$, $\delta_{J2000}$=$41^{\circ} 16^{\prime}
%08^{\prime\prime}.34$, \cite{Crane:1992}.)}
%\label{fig:super} 
%\end{figure*} 

\begin{figure*}
\centering
\includegraphics[width=6.6cm,angle=-90]{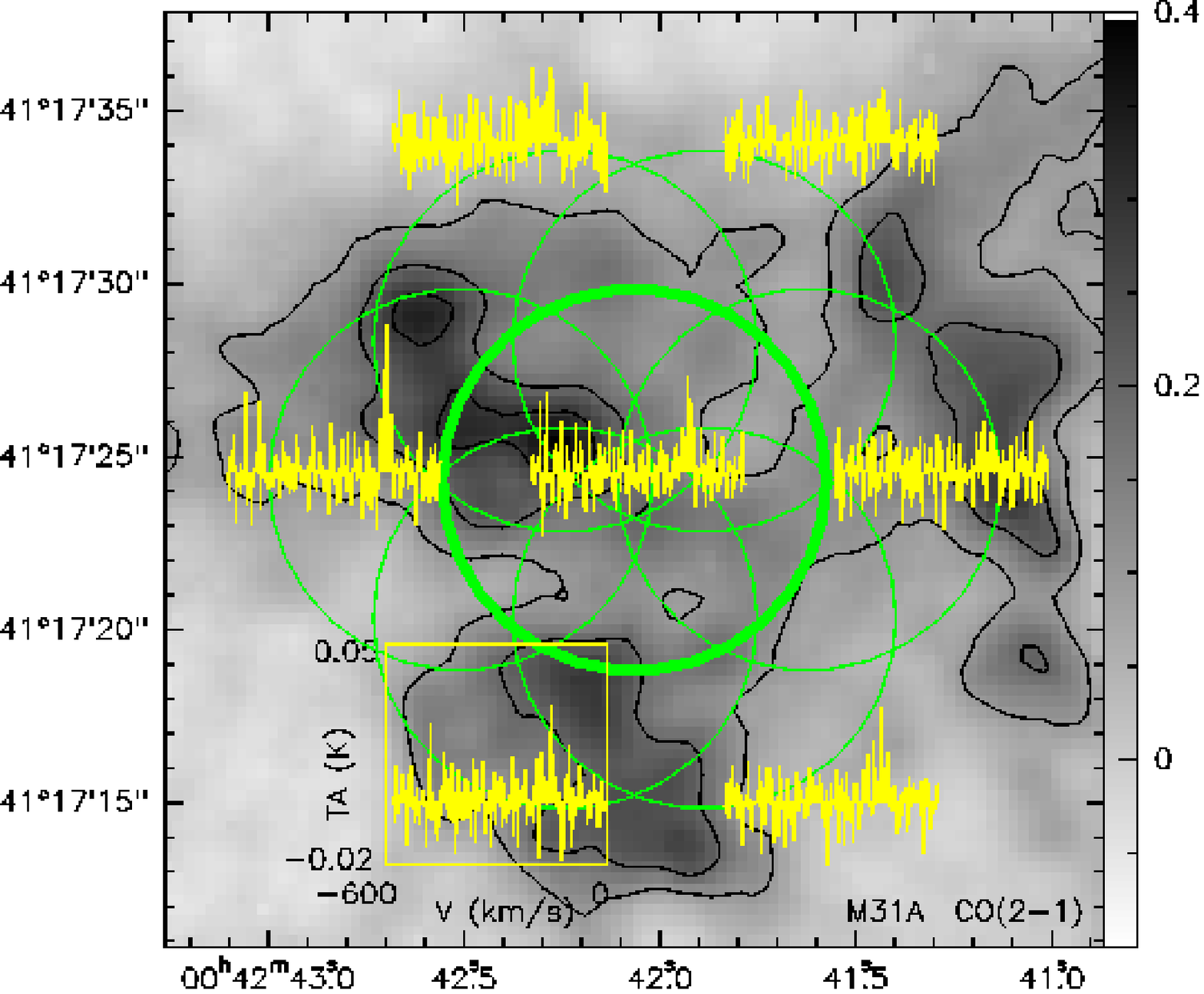}
\hspace{0.2cm}
\includegraphics[width=6.6cm,angle=-90]{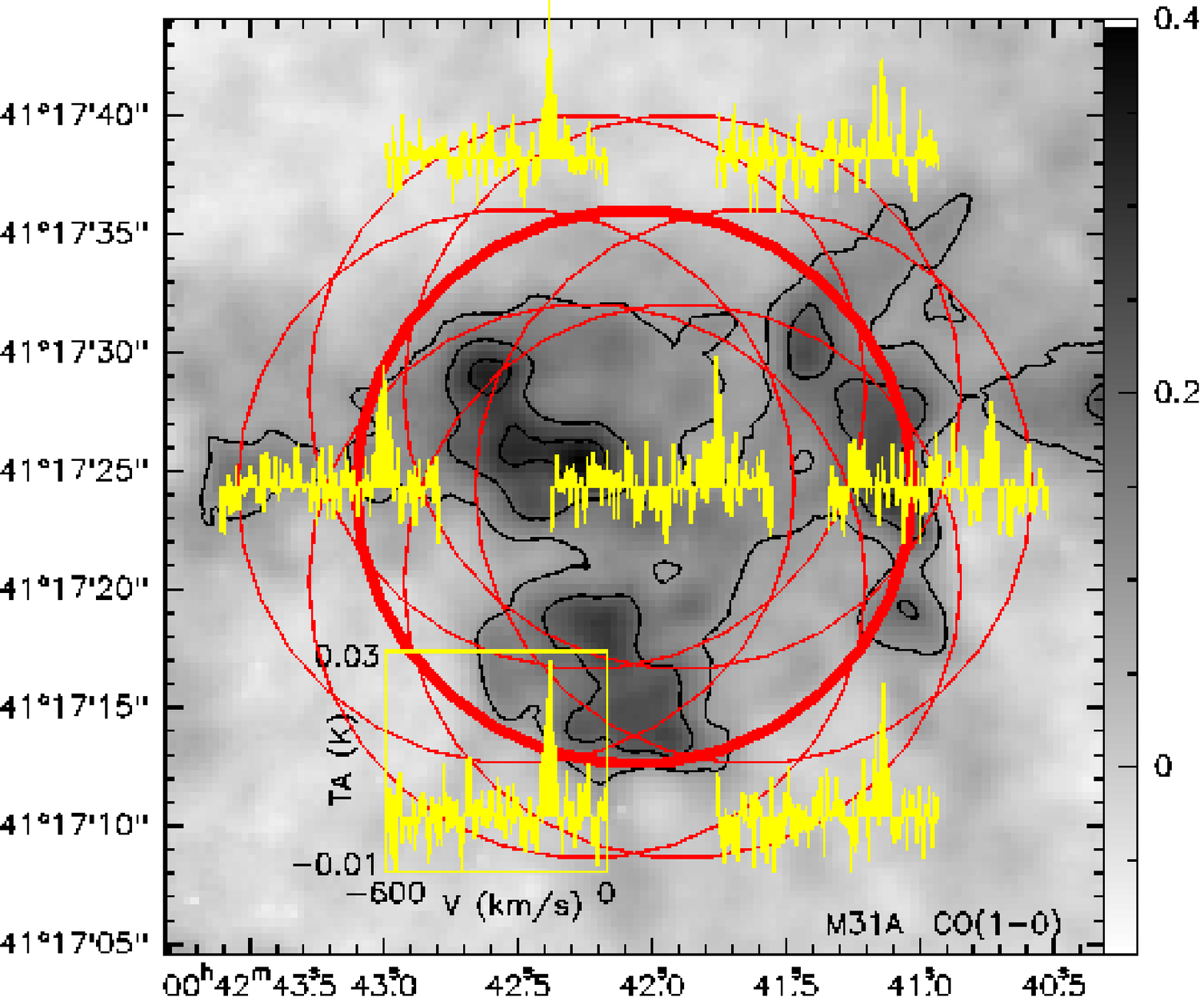}

\includegraphics[width=6.6cm,angle=-90]{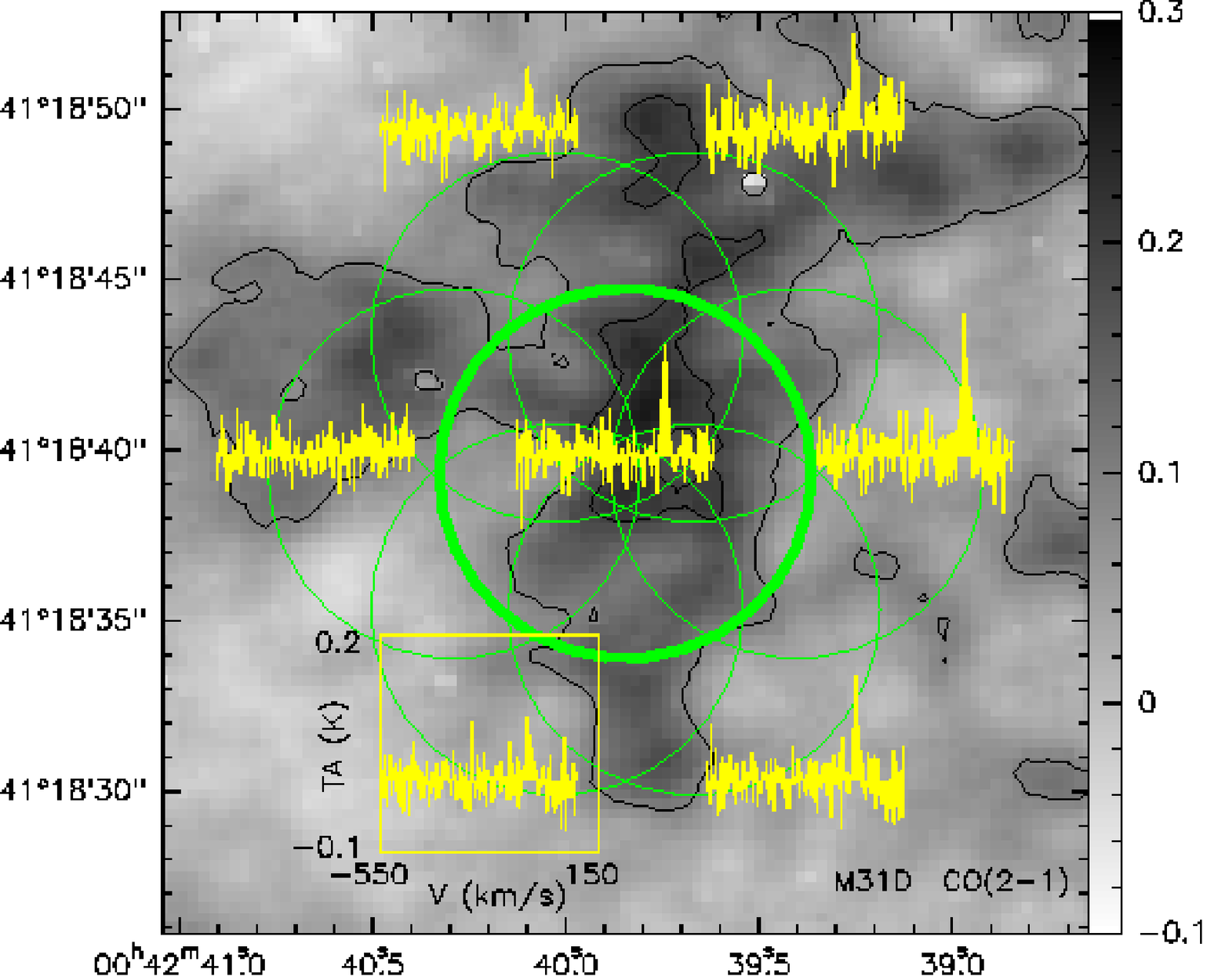}
\hspace{0.2cm}
\includegraphics[width=6.6cm,angle=-90]{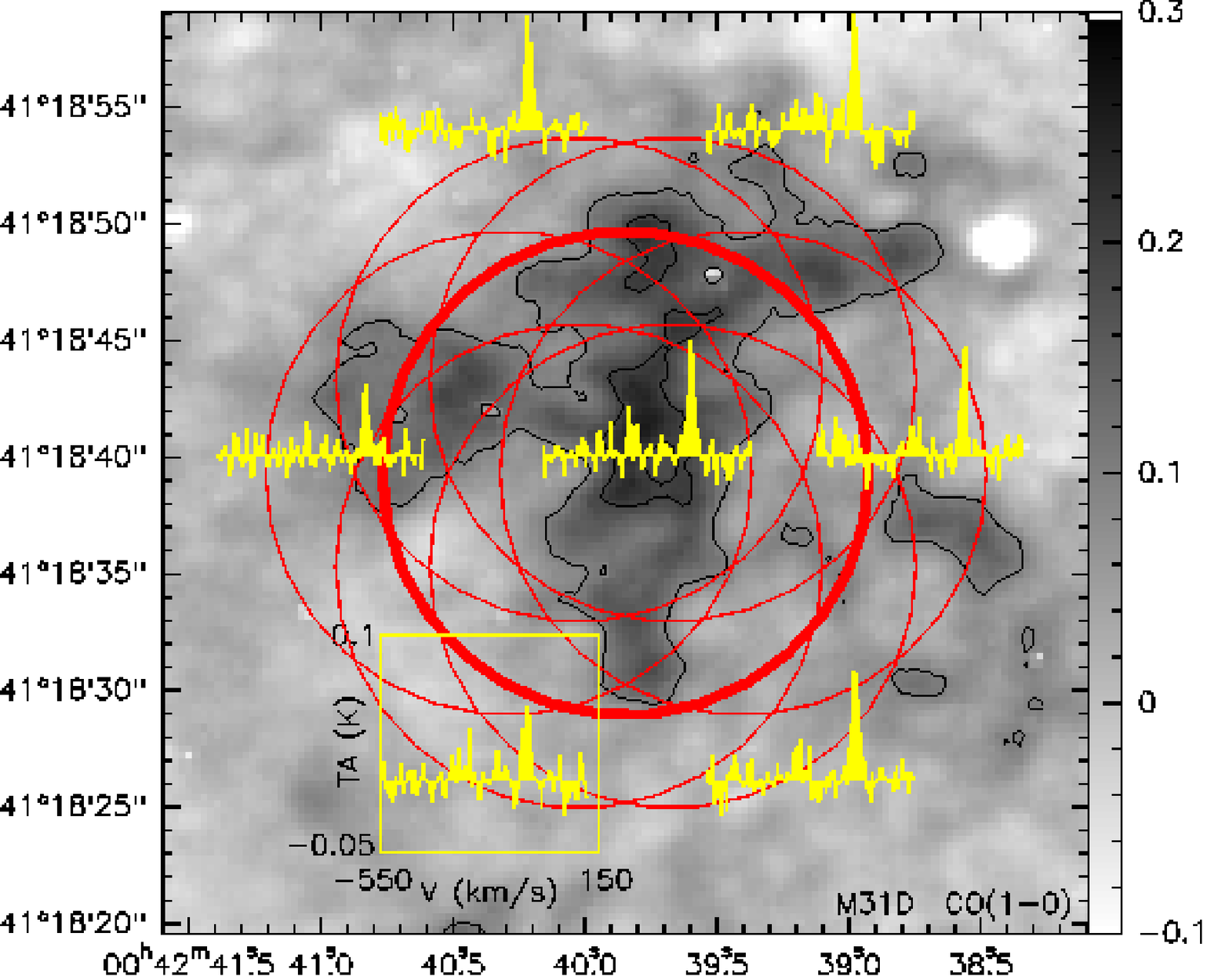}

\includegraphics[width=6.6cm,angle=-90]{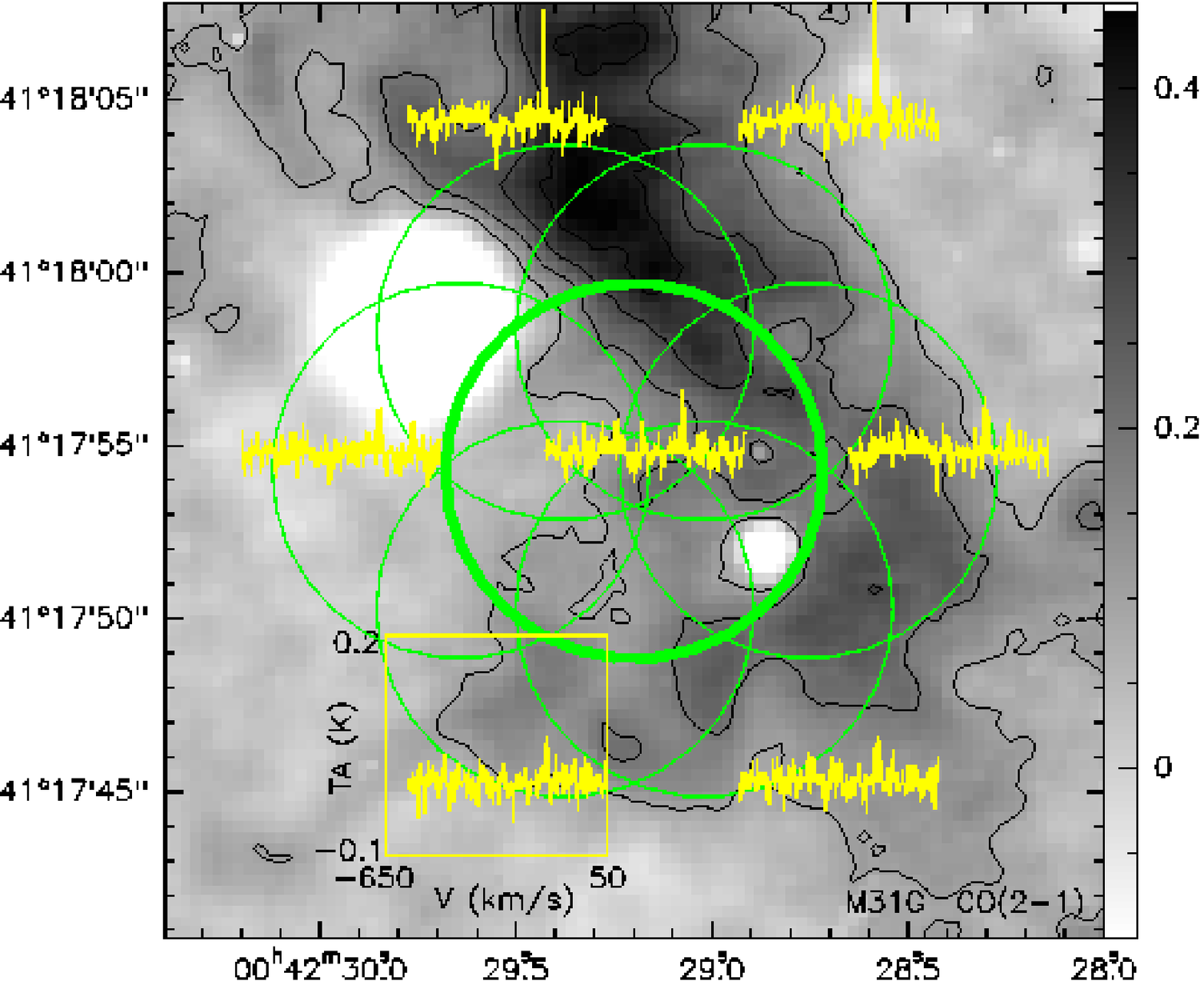}
\hspace{0.2cm}
\includegraphics[width=6.9cm,angle=-90]{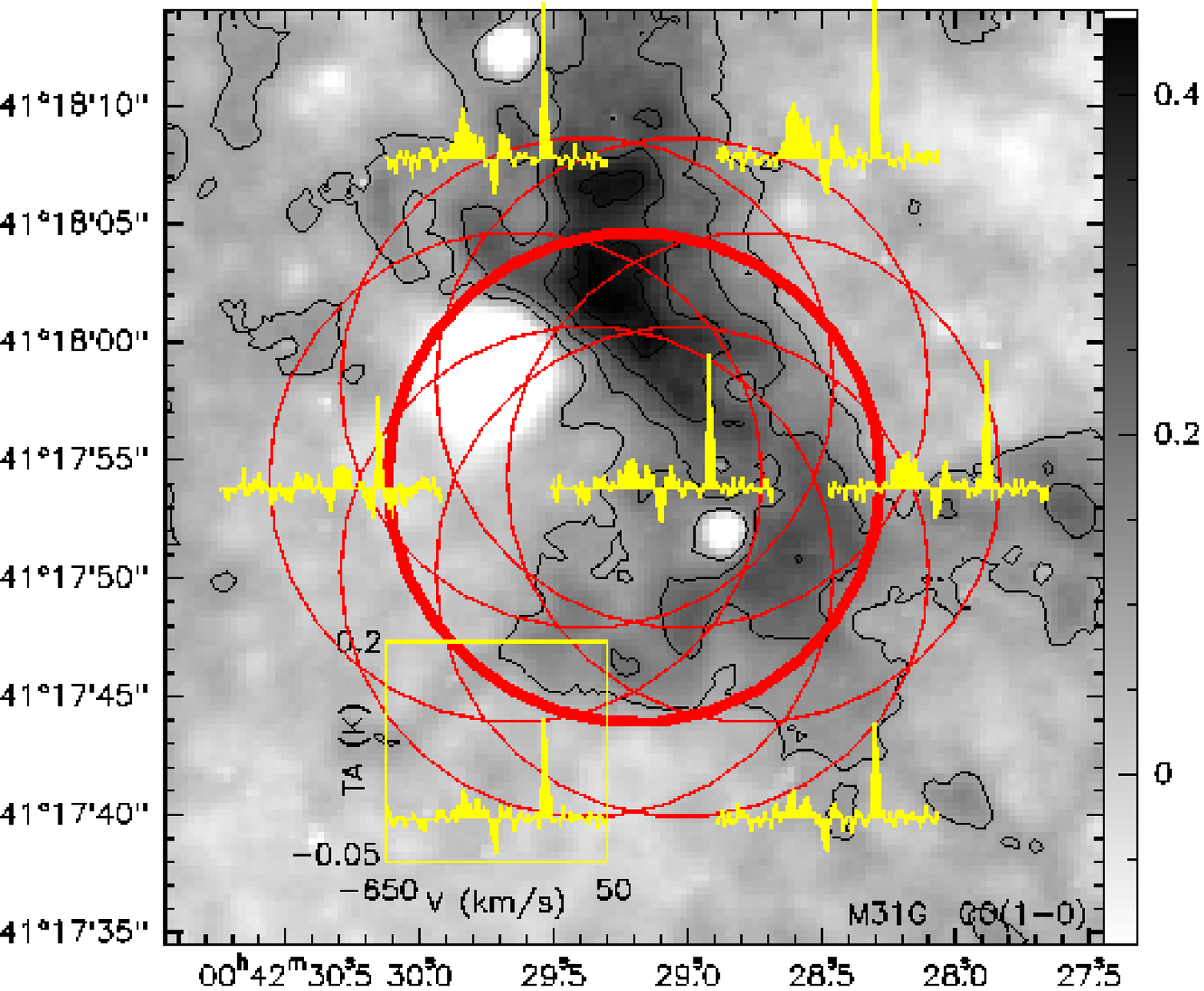}
\caption{Small maps performed at in CO(2-1) (left panels) and CO(1-0)
(right panels) for the positions A, D and G. The green (resp. red)
circles correspond to the FWHM of the CO(2-1) (resp. CO(1-0)) beams at
the various positions observed. The corresponding offsets are: (0,0),
(+5,0), (+2,+4) (-2,+4), (-5,0), (-2,-4), (+2, -4). The spectra are
superimposed on the corresponding A$_B$ maps. They are positioned
arbitrarily close to the beam corresponding to the observations.}
\label{fig:sub} % caption for the whole figure
\end{figure*}
The observations were carried out on 1999 June 13-15 and 2000 July
14-17 with the IRAM 30-m telescope. Most of the observations was made
in the symmetrical wobbler switching mode, in which the secondary
mirror nutates up to a maximum limit of $\pm$240 arcsec in
azimuth. The beam throw was determined as a function of the hour angle
in a way that the OFF positions lay in extinction-free regions, as
described in \cite{Melchior:2000}.  It nevertheless occurred in two
positions (M31D and M31G) that some signal was detected in the wobbler
throw positions.  As discussed in the Appendix \ref{sect:COOFF} and
indicated in Table \ref{tab:obs}, this corresponds to some azimuth
angles for M31G where one of the beam throws was close to the inner
spiral arm in the north/north-western part of the field. For M31D {
the configuration could be} more subtle and { the OFF signal} might
have been caused by gas located on the far side, which was not
detected in extinction.  In the reduction process, the OFF signal is
smeared out in the averaging process because equal weights are taken
to reduce the observations to an equivalent beam of 24\arcsec. It is
still noticeable in M31G because the OFF signal was also present in
the 2000 observations. Near transit we had to use position-switching
mode, taking an extinction-free OFF position located at a given
position from the nucleus as indicated in the last column of Table
\ref{tab:obs}.  We found the OFF signal for M31C for position-switch
observations, while no ON signal has been detected. These OFF
detections are presented in Appendix \ref{sect:COOFF}.

During the second epoch of observations small maps of seven points
with a spacing of ~5'' were made for positions (A, C, D and G) to
sample the CO(1-0) beam in CO(2-1) and to compute the CO(2-1)/CO(1-0)
line ratio. Pointing and focus calibration were regularly checked.  In
total, 10 ON positions in the inner disc of M\,31 were observed in
this way, as summarised in Table \ref{tab:obs} and presented in Figure
\ref{fig:super}. Table \ref{tab:obs} provides (1) the name of the
position observed, as used throughout the article, (2) the position
angle in degrees of this position, (3) the offset in arcsec of the ON
position with respect to the centre of M\,31, (4) the distance R of
this position with respect to the centre of M\,31, (5) the right
ascension, (6) the declination of this ON position, (7) the dates when
the observations were made, (8) the total integration time on the
source $T_{exp}$, (9) if a small map (seven pointings) was made
for this position, (10) if some signal was detected in the OFF
position, if yes, WSW (resp. PSW) indicates that some OFF signal was
detected in the wobbler (resp. position) switch scans performed for
this position, (11) the offsets in arcsec used for position-switch
observations made near transit.

%__________________________________________________ One column table
\begin{table*}
	\caption[]{Summary of observations}
	\label{tab:obs}
        \begin{tabular}{p{0.05\linewidth}p{0.05\linewidth}p{0.075\linewidth}p{0.05\linewidth}p{0.075\linewidth}p{0.075\linewidth}p{0.075\linewidth}p{0.025\linewidth}p{0.025\linewidth}p{0.025\linewidth}p{0.055\linewidth}}
            \hline
            \noalign{\smallskip}
            Position      &  PA & Offset &R  (arcsec)$~~$ & RA (J2000) & DEC (J2000) & date & T$_{exp}$ (min) & map & Off signal & Position switch (arsec)\\
            \noalign{\smallskip}
            \hline
            \noalign{\smallskip}
            M31A & 161.1 & -26,76 & 80.3 & 00:42:42.1 & 41:17:21.1 & 13/06/99     &  630 & no  & no  & 122,321 \\
                 & &           &   & &         & 14-17/07/00  & 1160 & yes & no  & 385,57; 274,23; 379,34\\
            M31C & -- & 0,0 & 0 & 00:42:44.4 & 41:16:09.2 & 2000         & 256  & yes & yes (PSW) & 385,57  \\
            M31D & 161.3 & -51,151 & 159.4 & 00:42:39.9 & 41:18:39.6 & 14/06/99     & 180  & no  & yes (WSW) & 122,321 \\
                 & &            &  & &          & 16/07/00     & 112  & yes & no  & 404,-41 \\
            M31E & 47.1 & 72,67 & 98.4 & 00:42:50.8 & 41:17:15.0 & 14/06/99     & 360  & no  & no  & 122,321 \\
            M31F & 125.6 & -134,96 & 164.8 & 00:42:32.5 & 41:17:44.0 & 15/06/99     & 442  & no  & no  & 122,321 \\
            M31G & 121.8 & -171, 106 & 201.2 & 00:42:29.2 & 41:17:54.0 & 15/06/99     & 180  & no  & yes (WSW) & 122,321 \\
               &  &            &    & &        & 15/07/00     & 175  & yes & yes (WSW)& 419,-7  \\
            M31GB&  170.7 & -22,134 & 135.8 & 00:42:42.4 & 41:18:22.0 & 15/06/99     & 78   & no  & no  & 122,321 \\
            M31I & 123.8 & -172,115 & 206.9 & 00:42:29.1 & 41:18:3.6  & 17/07/00     & 42   & no  & no  & 540,-23 \\
            M31J & 127.8 & -175,136 & 221.6 & 00:42:28.9 & 41:18:24.3 & 17/07/00     & 22   & no  & no  & 543,-43 \\
            M31K & 143.1 & -139,185& 231.4 & 00:42:32.0 & 41:19:13.7 & 17/07/00     & 90   & no  & no  & 508,-93 \\
            \noalign{\smallskip}
            \hline
         \end{tabular}
\end{table*}
We used four receivers simultaneously, two for $^{12}$CO(1-0) at
115\,GHz and two for $^{12}$CO(2-1) at 230\,GHz. At 115\,GHz each
receiver was connected to two autocorrelator sub-bands (shifted by
40\,MHz from each other) and each sub-band consisted of 225 channels
separated by 1.25\,MHz. At 230\,GHz each receiver was connected to a
filter-bank consisting of 512 channels of 1\,MHz width. 

\begin{table*}
	\caption[]{Characteristics of the CO lines. CO(2-1)
(resp. CO(2-1)) spectra are reduced to a 2.6\,\kms
(resp. 3.2\,\kms) resolution.}  
	\label{tab:lines}
        \begin{tabular}{p{0.1\linewidth}p{0.1\linewidth}p{0.12\linewidth}p{0.1\linewidth}p{0.1\linewidth}p{0.05\linewidth}p{0.05\linewidth}p{0.1\linewidth}p{0.05\linewidth}}
            \hline
            \noalign{\smallskip}
            Position      & $^{12}$CO line (beam size) & I$_{CO}$ (K km s$^{-1}$) $=\int T_{mb} dV$& V$_0$ (km s$^{-1}$) & $\sigma$ (km s$^{-1}$) & T$_{\rm peak}$ (mK) & baseline rms (mK) & $N_{H_2}$         (cm$^{-2}$) & $\Sigma_{H_2}$ (M$_\odot$\,pc$^{-2}$)\\
            \noalign{\smallskip}
            \hline
            \noalign{\smallskip}
            M31A    & 1-0 (24$\arcsec$) & 0.75$\pm$0.03 & -154.9$\pm$0.6 & 28.7$\pm$1.3 &  24.8 & 2.1& 1.73$\times 10^{20}$   &  2.94\\
                    & 2-1 (24$\arcsec$) & 0.88$\pm$0.06 & -151.1$\pm$0.7 & 22.9$\pm$1.9 &  35.7 & 4.7& $-$  		      &      \\
                    & 1-0 (21$\arcsec$) & 0.72$\pm$0.04 & -153.6$\pm$0.7 & 28.7$\pm$1.7 &  23.7 & 2.4& 1.67$\times 10^{20}$   &  2.83\\
                    & 2-1 (11$\arcsec$) & 0.78$\pm$0.06 & -152.8$\pm$1.0 & 24.9$\pm$2.6 &  29.4 & 5.5& $-$		      &      \\
            M31C    & 1-0 (24$\arcsec$) & $-$           & $-$            & $-$          &  $-$  & 3.7& $<$2.55$\times 10^{21}$& $<$43.31\\
                    & 2-1 (24$\arcsec$) & $-$           & $-$            & $-$          &  $-$  & 7.8& $-$		      &      \\
            M31D    & 1-0 (24$\arcsec$) & 2.43$\pm$0.10 &  -74.3$\pm$0.4 & 23.7$\pm$1.2 &  96.4 & 7.2& 5.58$\times 10^{20}$   &  9.49\\
                    & 2-1 (24$\arcsec$) & 3.45$\pm$0.16 &  -73.2$\pm$0.5 & 19.7$\pm$1.1 & 164.6 &15.5& $-$		      &      \\
                    & 1-0 (21$\arcsec$) & 2.47$\pm$0.11 &  -74.9$\pm$0.5 & 23.6$\pm$1.4 &  98.0 & 8.7& 5.68$\times 10^{20}$   &   9.66\\
                    & 2-1 (11$\arcsec$) & 2.12$\pm$0.12 &  -74.9$\pm$0.6 & 21.1$\pm$1.4 &  93.8 &10.6& $-$	      &       \\
M31E$^{\mathrm{a}}$ & 1-0 (21$\arcsec$) & 0.18$\pm$0.04 & -132.2$\pm$1.9 & 15.8$\pm$3.1 &  10.5 & 4.1& 4.25$\times 10^{19}$   &   0.72\\
                    &  2-1 (11$\arcsec$)& 0.33$\pm$0.10 & -128.2$\pm$3.1 & 15.3$\pm$6.9 &  19.4 &13.1& $-$		      &       \\
            M31F    & 1-0 (21$\arcsec$) & 1.56$\pm$0.14 & -360.6$\pm$3.3 & 68.0$\pm$6.7 &  20.6 & 6.4& 3.59$\times 10^{20}$   &   6.11\\
                    & 2-1 (11$\arcsec$) &  $-$          & $-$            & $-$          &   $-$ &16.3& $-$		      &       \\
            M31G    & 1-0 (24$\arcsec$) & 2.98$\pm$0.18 & -147.4$\pm$0.4 & 13.6$\pm$1.0 & 205.9 &4.9& 6.86$\times 10^{20}$   &  11.66\\
                    & 2-1 (24$\arcsec$) & 2.39$\pm$0.14 & -146.7$\pm$0.5 & 15.4$\pm$1.1 & 145.2 &9.8& $-$		      &       \\
                    & 1-0 (21$\arcsec$) & 3.01$\pm$0.23 & -146.7$\pm$0.6 & 15.3$\pm$1.4 & 185.2 &5.3& 6.92$\times 10^{20}$   &  11.77\\
                    & 2-1 (11$\arcsec$) & 2.18$\pm$0.20 & -146.6$\pm$0.9 & 18.5$\pm$1.9 & 110.6 &15.9& $-$		      &       \\
            M31GB   & 1-0 (21$\arcsec$) & $-$           & $-$            & $-$          &   $-$ & 7.7& $<$5.29$\times 10^{21}$&  $<$89.95\\
                    & 2-1 (11$\arcsec$) & $-$           & $-$            & $-$          &   $-$ &14.7& $-$                    &      \\		        
            M31I    & 1-0 (21$\arcsec$) & 3.22$\pm$0.11 & -148.5$\pm$0.2 &  9.3$\pm$0.4 & 324.3 &12.5& 7.41$\times 10^{20}$   &  12.60\\
		    & 1-0 (21$\arcsec$) & 4.60$\pm$0.77 & -411.9$\pm$1.0 & 34.1$\pm$3.1 & 126.8 &12.5& 1.06$\times 10^{21}$   &  17.99\\
		    & 1-0 (21$\arcsec$) & 2.50$\pm$0.31 & -278.7$\pm$2.6 & 46.3$\pm$5.4 &  50.7 &12.5& 5.75$\times 10^{20}$   &   9.77\\
		    & 1-0 (21$\arcsec$) & 3.69$\pm$0.89 & -376.4$\pm$11.1& 88.5$\pm$16.5&  39.2 &12.5& 8.49$\times 10^{20}$   &  14.44\\
                    & 2-1 (11$\arcsec$) & 3.28$\pm$0.12 & -148.6$\pm$0.2 &  7.7$\pm$0.4 & 400.9 &29.0& $-$                    &       \\
                    & 2-1 (11$\arcsec$) & 7.28$\pm$0.47 & -409.5$\pm$1.3 & 40.0$\pm$3.5 & 171.0 &29.0& $-$                    &       \\
                    & 2-1 (11$\arcsec$) & 4.04$\pm$0.47 & -283.4$\pm$2.7 & 43.5$\pm$5.0 &  87.7 &29.0& $-$                    &       \\
                    & 2-1 (11$\arcsec$) & 2.96$\pm$0.41 & -353.3$\pm$5.7 & 46.6$\pm$4.5 &  59.6 &29.0& $-$                    &       \\
            M31J    & 1-0 (21$\arcsec$) &12.81$\pm$0.33 & -413.2$\pm$0.2 & 33.3$\pm$0.8 & 361.4 &12.2& 2.95$\times 10^{21}$   &  50.08\\
		    & 1-0 (21$\arcsec$) & 1.06$\pm$0.11 & -145.5$\pm$0.6 & 11.9$\pm$1.5 &  83.8 &12.2& 2.45$\times 10^{20}$   &   4.16\\
		    & 1-0 (21$\arcsec$) & 7.34$\pm$0.57 & -350.1$\pm$7.5 &178.2$\pm$13.2&  38.8 &12.2& 1.69$\times 10^{21}$   &  28.70\\
                    & 2-1 (11$\arcsec$) &15.22$\pm$0.78 & -412.7$\pm$0.9 & 35.1$\pm$2.0 & 407.0 &54.5& $-$                    &       \\
                    & 2-1 (11$\arcsec$) & 0.86$\pm$0.41 & -142.1$\pm$3.0 &  9.0$\pm$5.2 &  88.5 &54.5& $-$                    &       \\
                    & 2-1 (11$\arcsec$) & 1.94$\pm$0.69 & -268.6$\pm$7.1 & 35.4$\pm$10.8&  51.4 &54.5& $-$                    &       \\
            M31K    & 1-0 (21$\arcsec$) & 2.66$\pm$0.14 & -300.8$\pm$1.1 & 41.7$\pm$2.9 &  59.8 & 7.5& 6.11$\times 10^{20}$   &  10.38\\
		    & 2-1 (11$\arcsec$) & 2.61$\pm$0.43 & -292.5$\pm$4.9 & 58.5$\pm$11.0&  42.0 &24.1& $-$                    &        \\
            \noalign{\smallskip}
            \hline
         \end{tabular}
\begin{list}{}{}
\item[$^{\mathrm{a}}$] Tentative detection.
\end{list}
\end{table*}

\section{Data reduction} 
\label{sect:datared}
The {\sc{CLASS}}\footnote{Continuum and Line Analysis Single-dis
Software, http://www.iram.fr/IRAMFR/GILDAS} package was used for
the data reduction. After checking the quality of each single spectra,
the data were averaged with inverse variance weights. A
first-order baseline was fitted to the resulting spectrum and
subtracted. For a few spectra, a higher-order polynomial was
carefully subtracted. Finally, the spectra were smoothed to a
velocity resolution of 3.2\,\kms (resp. 2.6\,\kms) in
CO(1-0) (resp. CO(2-1)).

The maps are presented in Figure \ref{fig:sub} and illustrate the
observing procedure. The spectra are located arbitrarily close to the
beam position they correspond to.  Spatial variations in the spectra
agree relatively well with the extinction distribution. For M31G the
CO intensity is clearly stronger in the area where the extinction is
stronger. For M31A and M31D the intensity of the lines does not follow
the intensity of the extinction exactly but we cannot exclude that
some pointing errors and/or a clumpy distribution of the underlying
gas explains the spatial variations of the CO distribution.

The results of the fitting are presented in Table \ref{tab:lines}. A
Gaussian function is fitted to each line to determine its area,
central velocity $V_0$, width $\sigma$, and peak temperature $T_{\rm
peak}$. The baseline RMS is provided for each line. We provide
main-beam temperatures (unless specified otherwise) throughout this
paper with B$_{eff}=$ 64.2$\pm 3$ and F$_{eff}=$ 91$\pm 2$
(resp. B$_{eff}=$ 42.$\pm 3$ and F$_{eff}=$ 86$\pm 2$) at 115\,GHz
(resp. 230\,GHz). For the positions for which maps have been observed,
the spectra were convolved to obtain an 24$\arcsec$ beam ($\sqrt{11^2
+ 21^2}$) for both lines, enabling a direct computation of the line
ratio.

\begin{figure*}
\centering 
   \includegraphics[width=6.3cm,angle=-90]{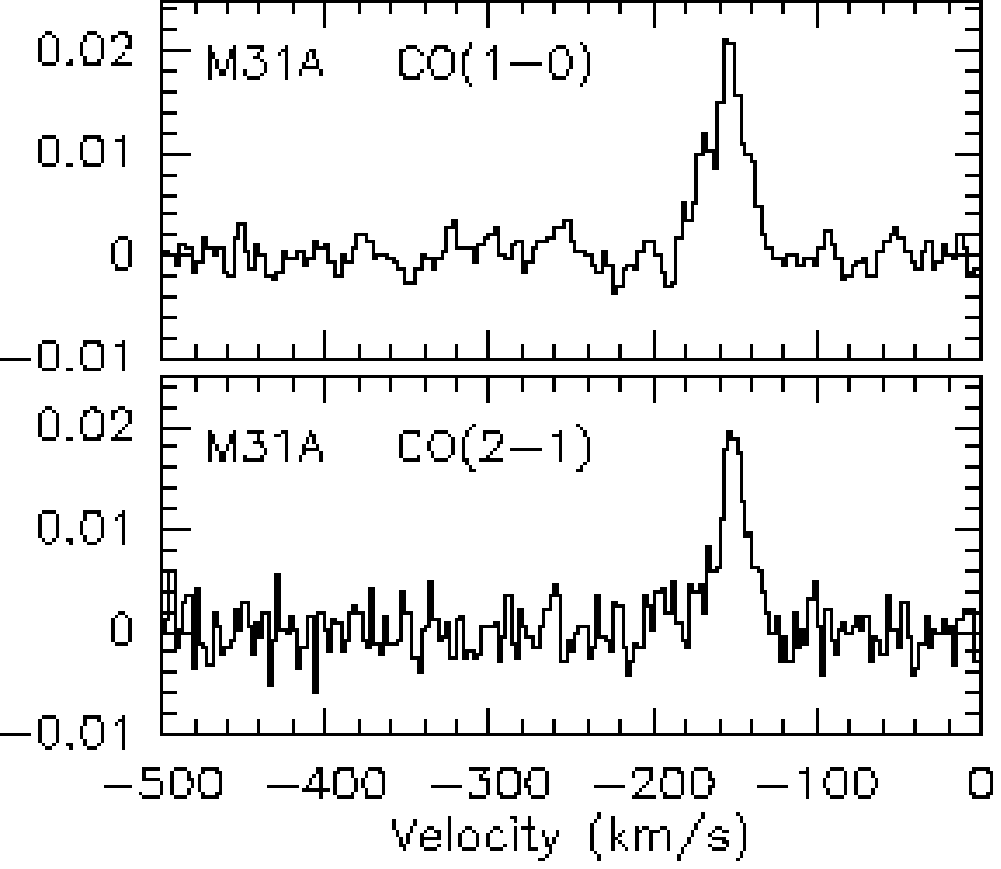}
   \includegraphics[width=6.3cm,angle=-90]{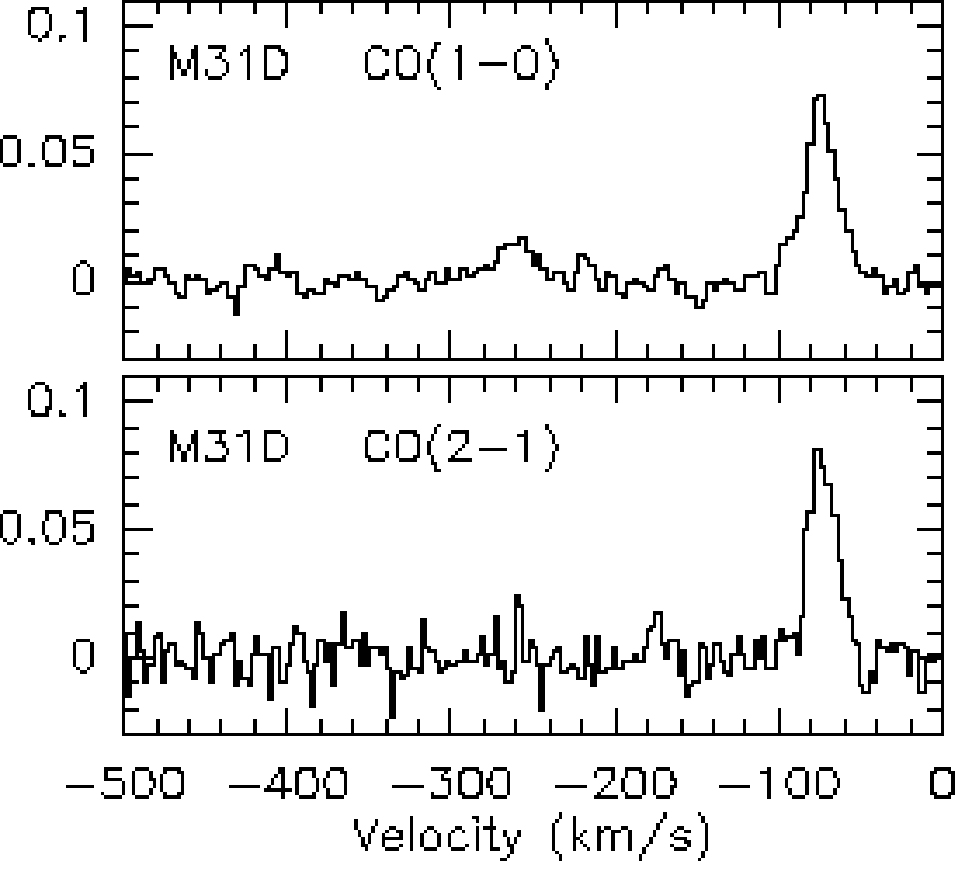}
   \includegraphics[width=6.3cm,angle=-90]{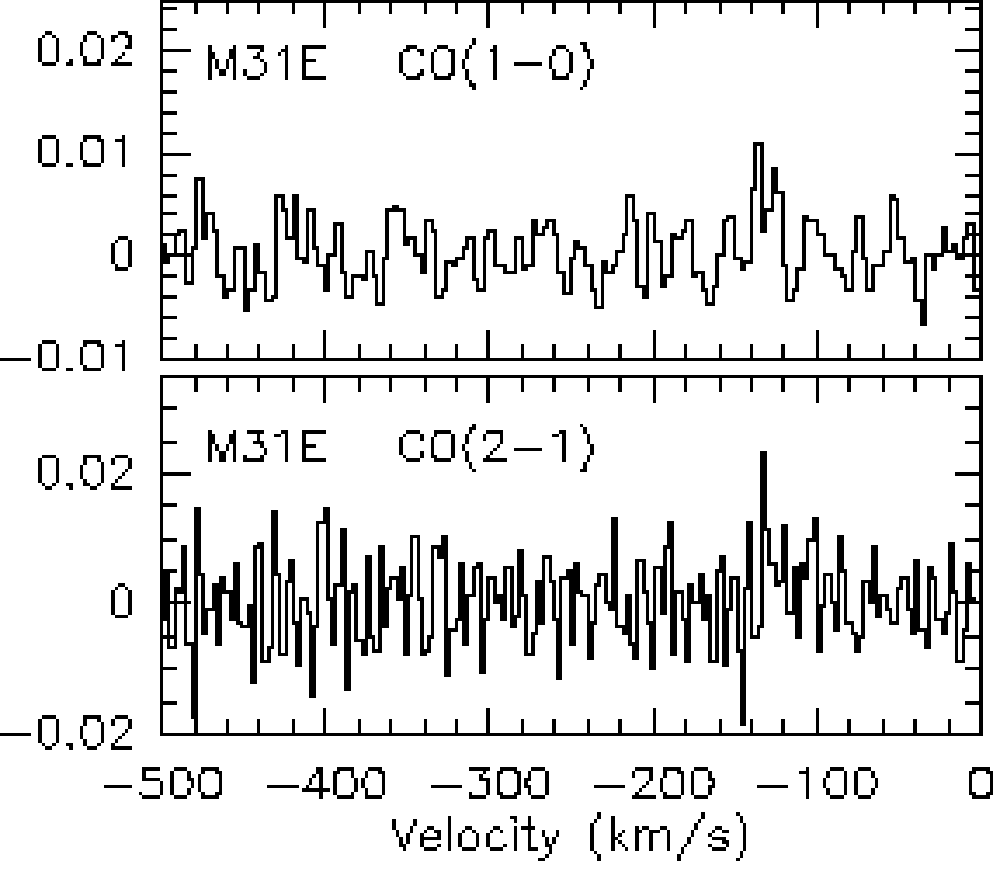}
   \includegraphics[width=6.3cm,angle=-90]{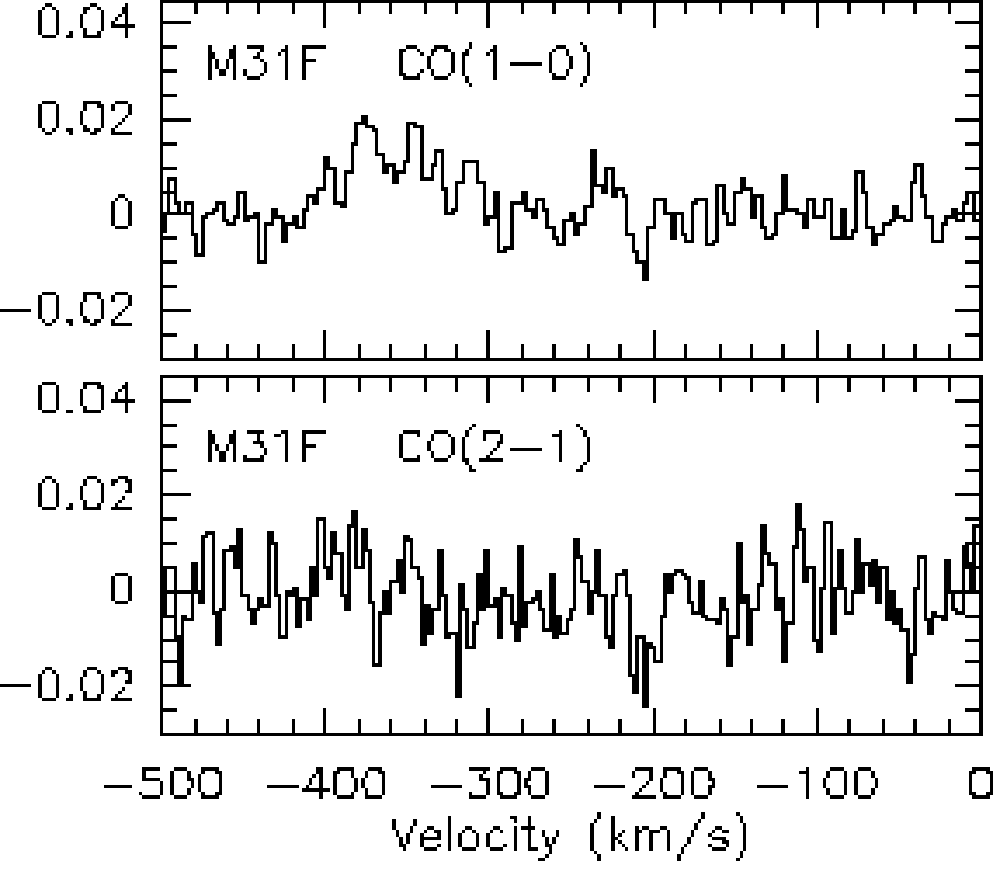}
   \includegraphics[width=6.3cm,angle=-90]{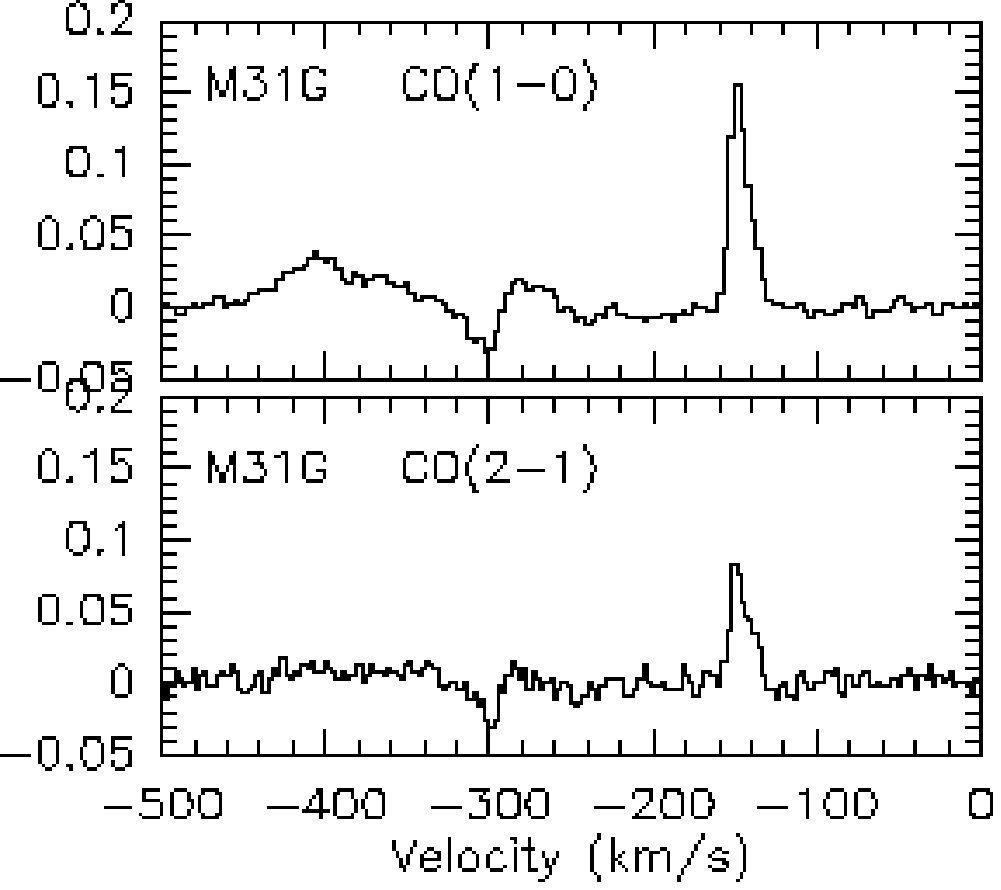}
   \includegraphics[width=6.3cm,angle=-90]{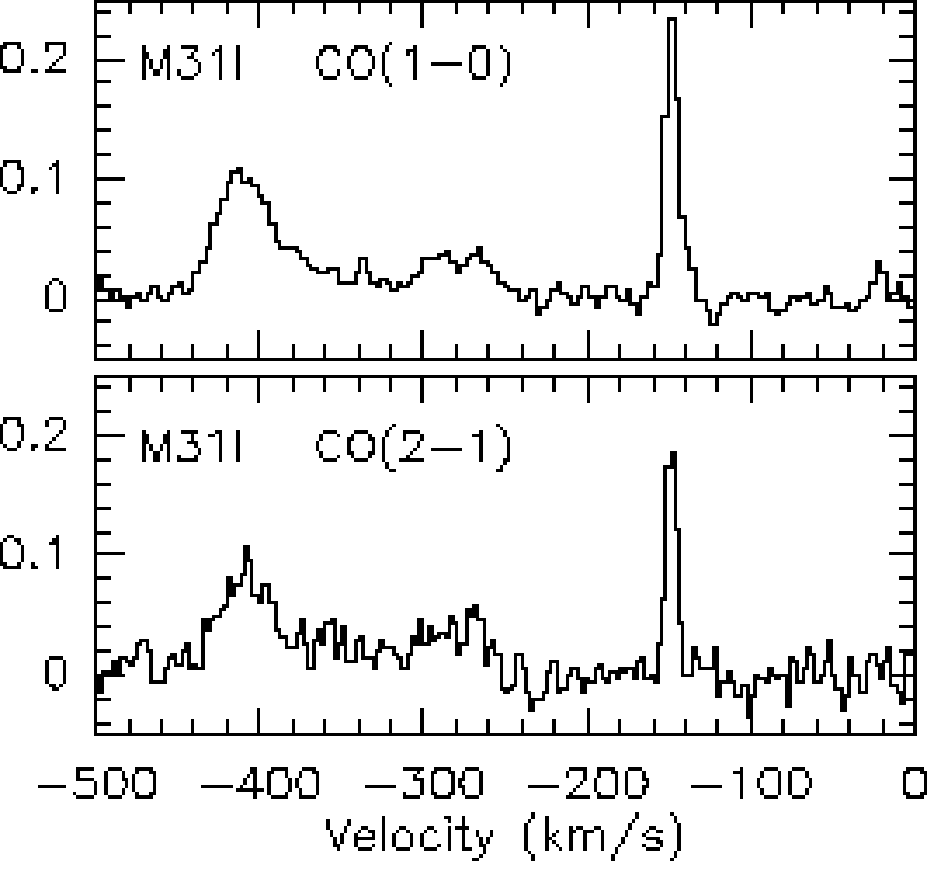}
   \includegraphics[width=6.3cm,angle=-90]{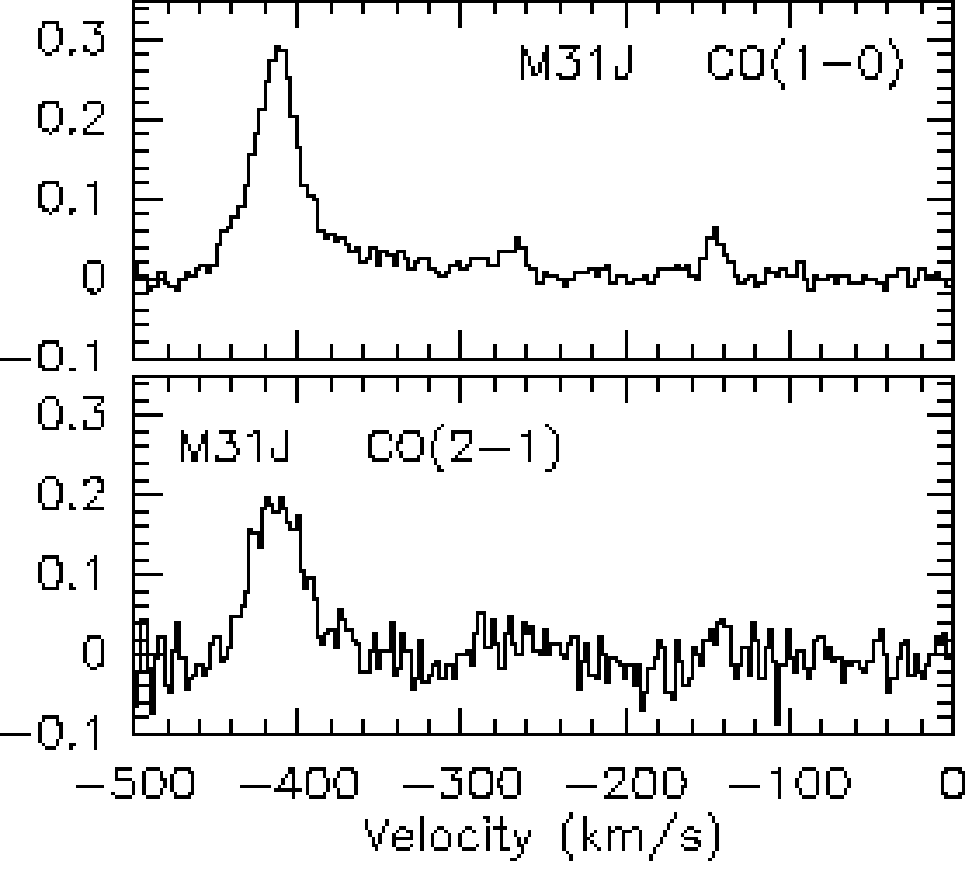}
   \includegraphics[width=6.3cm,angle=-90]{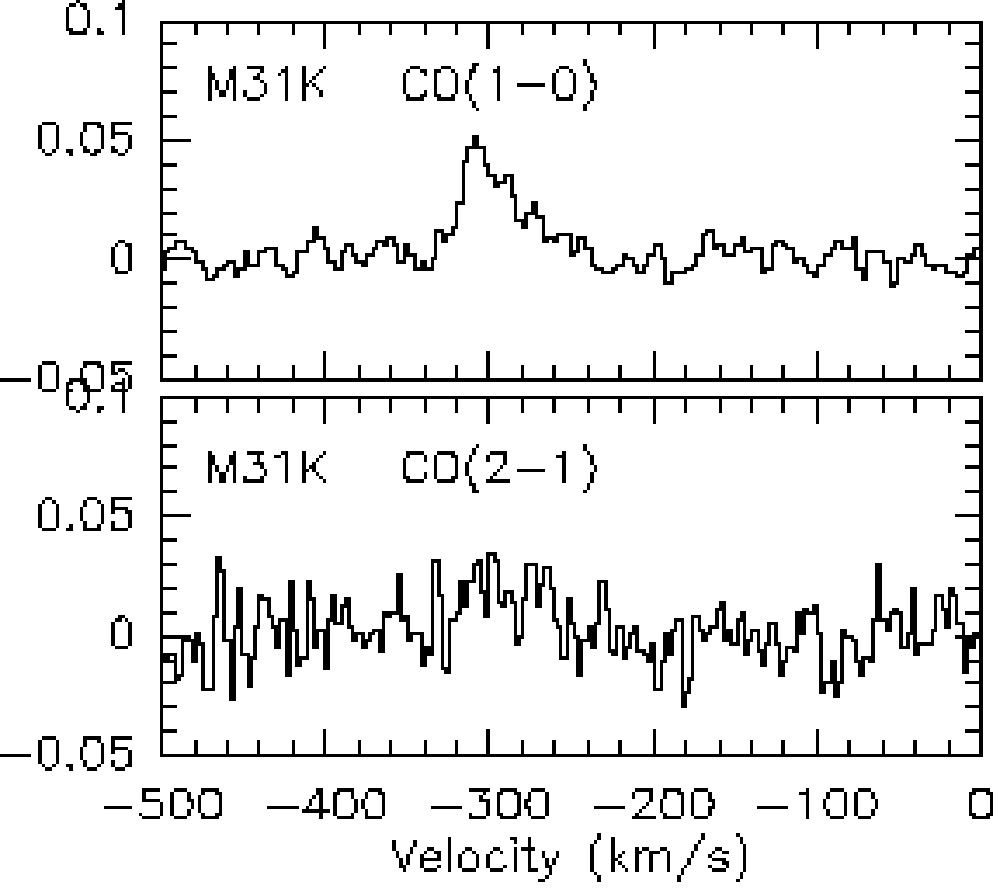} \caption{Spectra
   with a detected signal. M31A, M31D and M31G (resp. M31E, M31F, M31I
   and M31J) spectra are reduced to a 24$\arcsec$ beam (resp. with the
   original beam) in CO(2-1) and CO(1-0). The y-axis provides the
   antenna temperature in K.} \label{fig:M31all}
\end{figure*}

The spectra are displayed in Figure \ref{fig:M31all}. In the final
spectra, the signal present in the OFF positions is mainly visible for
the position M31G. It is at the systemic velocity and probably
corresponds to the gas in the main disc along the minor axis in the
north of the M31G position (because the OFF positions are symmetric in
azimuth with respect to the ON position). { We use probably throughout
this paper except for the \citet{Saglia:2010} data, whose velocities
are formally in LSR, but the difference ($\sim$5km\,s$^{-1}$) is
negligible here.}

%----------------------------------------------------------- 
%
%______________________________________________________________

\section{Analysis}
\label{sect:ana}
\begin{table*}
	\caption[]{A$_B$ extinction derived from the map based on the
	\citet{Ciardullo:1988} data as obtained in
	\citet{Melchior:2000}. For each configuration E(B-V) was then
	computed assuming R$_V$ in the range $[2.1,3]$. These
	values were computed assuming a fraction of foreground
	light $x=0$. The I$_{CO}/E(B-V)$ ratio is then provided for
	the strongest lines. For M31J we also provide the ratio
	corresponding to the ring velocity.}
	\label{tab:abs}
        \begin{tabular}{llllllll}
            \hline
            \noalign{\smallskip}
Position     & A$_B$ at 24$\arcsec$ & A$_B$ at 21$\arcsec$ (11$\arcsec$) & 
	E(B-V) at 24$\arcsec$ &  E(B-V) at 21$\arcsec$ &  E(B-V) at
11$\arcsec$  & I$_{CO}$/E(B-V) at 24$\arcsec$ &I$_{CO}$/E(B-V) at 21$\arcsec$ \\
& (mag beam$^{-1}$) & (mag beam$^{-1}$)   & (mag
beam$^{-1}$)  & (mag beam$^{-1}$)  & (mag beam$^{-1}$) & (K km s$^{-1}$
mag$^{-1}$)& (K km s$^{-1}$
mag$^{-1}$)\\
            \noalign{\smallskip}
            \hline
            \noalign{\smallskip}
M31A & 0.095 & 0.100 (0.166) & 0.027$\pm$0.003 &0.029$\pm$0.004 &0.048$\pm$0.006& 28$\pm$5 & 26$\pm$ 5\\
M31D & 0.067 & 0.076 (0.111) & 0.019$\pm$0.002 &0.022$\pm$0.003 &0.032$\pm$0.004&129$\pm$22&116$\pm$20\\
M31F & 0.074 & 0.081 (0.123) & 0.021$\pm$0.003 &0.023$\pm$0.003 &0.035$\pm$0.004&          & 69$\pm$15\\
M31G & 0.182 & 0.193 (0.256) & 0.052$\pm$0.007 &0.055$\pm$0.007 &0.073$\pm$0.009& 59$\pm$11& 56$\pm$11\\
M31I & 0.210 & 0.219 (0.245) & 0.060$\pm$0.008 &0.063$\pm$0.008 &0.070$\pm$0.009&          & 52$\pm$ 8\\
M31J & 0.203 & 0.208 (0.241) & 0.058$\pm$0.007 &0.060$\pm$0.008 &0.069$\pm$0.009&          & 219$\pm$33(18$\pm$ 4)\\
M31K & 0.138 & 0.141 (0.159) & 0.040$\pm$0.005 &0.040$\pm$0.005 &0.046$\pm$0.006&          & 67$\pm$12\\
            \noalign{\smallskip}
            \hline
         \end{tabular}
\end{table*}
\subsection{Characteristics of the CO emission and gas-dust connection}
The CO gas, which we searched for and discuss in this paper, is
located mainly around the minor axis of the main disc of M31. The
detected molecular gas complexes correspond to the strongest A$_B$
extinction complexes that we observed. For the positions without
extinction (M31GB), we detected no signal, while no
line was detected in the centre either. For M31E where the
extinction is weak, we have a tentative detection at 4$\sigma$.  The
extinction wascomputed assuming a fraction of foreground light
of $x=0$ \citep[see][]{Melchior:2000}, which is probably correct for
the north-western part, where we observed, because an inclination of
45\,$\deg$ is usually assumed for the nuclear spiral
\citep{Ciardullo:1988}.  {We also computed the near-UV and far-UV
extinction maps from archive GALEX data with the same method and found
the same structures. In addition, we subtracted from the original
GALEX images the bulge emission modelled in this way, and found no
trace of UV emission in the inner ring.}

   \begin{figure*} \centering
   \includegraphics[height=14cm,angle=-90]{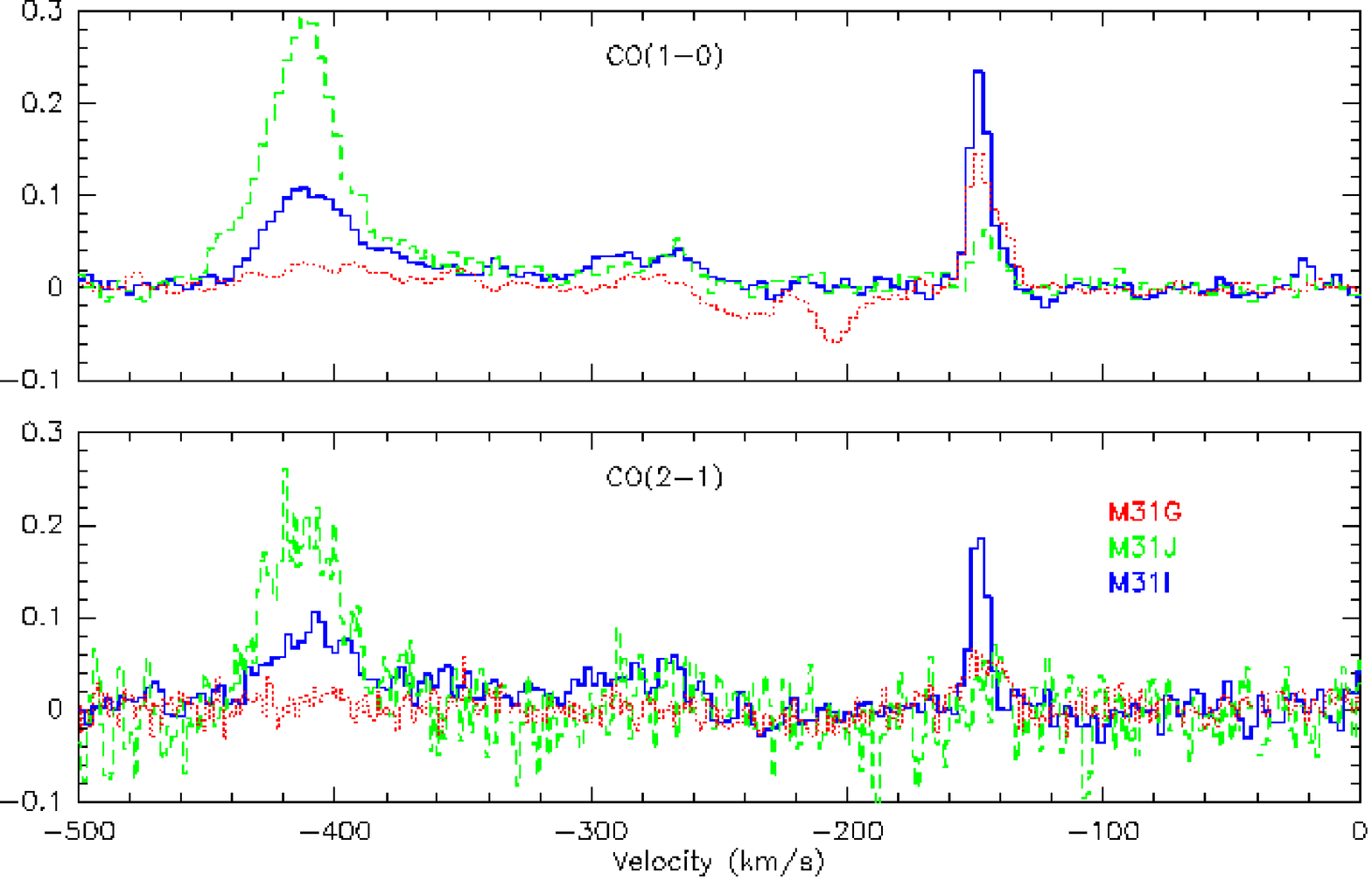}
   \caption{Superimposition of the G, I, and J spectra at the nominal
   spatial resolution, namely 21$\arcsec$ for CO(1-0) and 11$\arcsec$
   for CO(2-1).  The y-axis provides the
   antenna temperature in K.} \label{fig:superimp} \end{figure*}

The position M31F, which exhibits a weak CO(1-0) signal with no
detection in CO(2-1), corresponds to a position with low A$_B$
extinction. The other positions are detected with high signal-to-noise
ratios at least in CO(1-0). We computed the A$_B$ extinction values
for each position convolved with the different beams, as provided in
Table \ref{tab:abs}, and there is no one-to-one correspondence with
the intensity of the detected CO. Figure \ref{fig:super} also shows
that M31D has a much stronger signal in CO than M31A while its A$_B$
extinction is weaker. One plausible explanation could be that the dust
clumps corresponding to M31D do not lie in the same plane as M31A: if
the foreground light ($x$) exceeds 0, we have underestimated the
extinction. Assuming the I$_{CO}$/A$_B$ ratio measured for M31A
applies to M31D, one would expect a peak intensity A$_B^{real} = 1.56$
for M31D, but it is measured A$_B^{meas.} = 0.26$. This configuration
corresponds to a fraction of light in front of the dust $x=0.72$,
meaning that M31D lies on the back side of the bulge. It is also
probable that the gas is very clumpy and that the non-linearity of the
extinction somehow biases our A$_B$ estimate.

Our detections are concentrated in an area\footnote{As in
\citet{Melchior:2000}, we assume a distance of M\,31 of 780\,kpc
i.e. 1\,arcsec$=$3.8\,pc.} of 415\,pc $\times$ 570\,pc (in
projection), located in the north-western part of the M\,31 within
3.8\,arcmin from the centre, corresponding to 880\,pc (resp.  1.2\,kpc
if deprojected {for a 45$\deg$ inclination}).  While the area explored
is quite localised, the detected velocities span from -73 to
-413\,\kms, while velocities along the minor axis of a rotating disc
are expected to be at the systemic velocity. M31K is the only position
exhibiting this expected velocity, as well as a component for M31G,
M31I and M31J, and possibly M31D. Figure \ref{fig:superimp} shows a
superimposition of the spectra obtained for M31G, M31I and M31J (and
reduced to a 24\arcsec\, beam): the various velocity patterns appear
in each spectra with different relative intensities. Besides the
velocity line detected at -145\,\kms, a broadband signal is detected
with velocities between -450 and -250\,\kms. As discussed below, this
range of velocities cannot be explained by a single disc inclined at
45\,deg in regular rotation, and they are the signature of a peculiar
structure. The line widths of the detected lines are quite different
from one location to another, suggesting a spatial extension of the
emitting areas. We assume a Galactic $X_{CO} = N_{H_2}/I_{CO} = 2.3
\times 10^{20}$\,cm$^{-2}$(K\,\kms)$^{-1}$ following
\citet{Strong:1988}. For the positions where we did not detect any
signal, we provide a 3\,$\sigma$ upper limit based on the
dispersion (rms) computed on the baseline.  Some CO(2-1) measurements
(M31I, M31J, M31K) are at the limit of detection and are provided for
a rough comparison with CO(1-0) measurements.

In the centre {($<2\arcsec$)}, our $A_B$ map does not measure the
extinction (owing to the method, which is based on ellipse fitting).
The map of 8\,$\mu$m emission (after subtraction of the scaled stellar
continuum at 3.6\,$\mu$m) also displays a defect in the central part,
so it is difficult to be conclusive about the extinction in the inner
arcseconds. However, \citet{Garcia:2000}, relying on simple modelling
of {\em Chandra} data, have estimated $A_V=1.5\pm 0.6$ in the central
3-arcsec region. Our millimetre observations exclude the presence of
CO(1-0) (resp. CO(2-1)) at the 1\,$\sigma$ level of 3.7\,mK
(resp. 7.8\,mK). On the basis of this upper limit, the gas present
within 80\,pc from M31's centre does not exceed
43\,M$_\odot$/pc$^{-2}$ (3\,$\sigma$), assuming a conservative value
for X$_{CO}$ typical of the Galactic disc.  This is an extremely small
amount compared to the gas mass computed in the central region of the
Milky Way \citep{Oka:1998}.
\begin{table}
	\caption[]{Line ratios of the complexes reduced to a 24$\arcsec$
beam. We provide both the ratios of the line intensities r$_{12}$ and the
ratios of the peak temperatures $R$.}  
	\label{tab:linerat}
        \begin{tabular}{lll}
            \hline
            \noalign{\smallskip}
            Complexes & r$_{12}$ & $R$   \\
            \noalign{\smallskip}
            \hline
            \noalign{\smallskip}
M31A & 1.17$\pm^{0.40}_{0.19}$ & 1.44$\pm^{0.3}_{0.1}$  \\
M31D & 1.43$\pm^{0.45}_{0.21}$ & 1.71$\pm^{0.36}_{0.12}$  \\
M31G & 0.80$\pm^{0.30}_{0.13}$ & 0.71$\pm^{0.14}_{0.05}$  \\
            \noalign{\smallskip}
            \hline
         \end{tabular}
\end{table}

We used the three positions with observations reduced to a 24\arcsec\,
beam to compute the CO(2-1) to CO(1-0) line ratio. As displayed in
Table \ref{tab:linerat}, we computed both the ratio of the integrated
intensities and the ratio of the peak temperatures, which are
compatible within error bars. M31G exhibits a line ratio comparable
with the mean value found for the Milky Way spiral arms
\citep{Sakamoto:1997}.

M31A and M31D exhibits CO line ratios higher than 1.  Given the
N$_{H_2}$ column densities computed (respectively $5.6\times 10^{20}$
cm$^{-2}$ and $1.7\times 10^{20}$ cm$^{-2}$), the gas is probably not
optically thin.  In addition, the velocity dispersion of CO(2-1) is
systematically smaller than that of CO(1-0), suggesting that these
complexes are probably externally heated and composed of dense
gas. This hypothesis is compatible, as discussed below, with the
combined presence of an intense stellar radiation field caused by
bulge stars \citep{Stephens:2003}, the outflow detected in X-ray in
this area \citep{Bogdan:2008}, and the ionised gas
\citep{Jacoby:1985}. This supports the view that no star formation is
currently happening inside the inner ring of M31.
\begin{figure*}
\centering 
   \includegraphics[width=6.5cm,angle=-90]{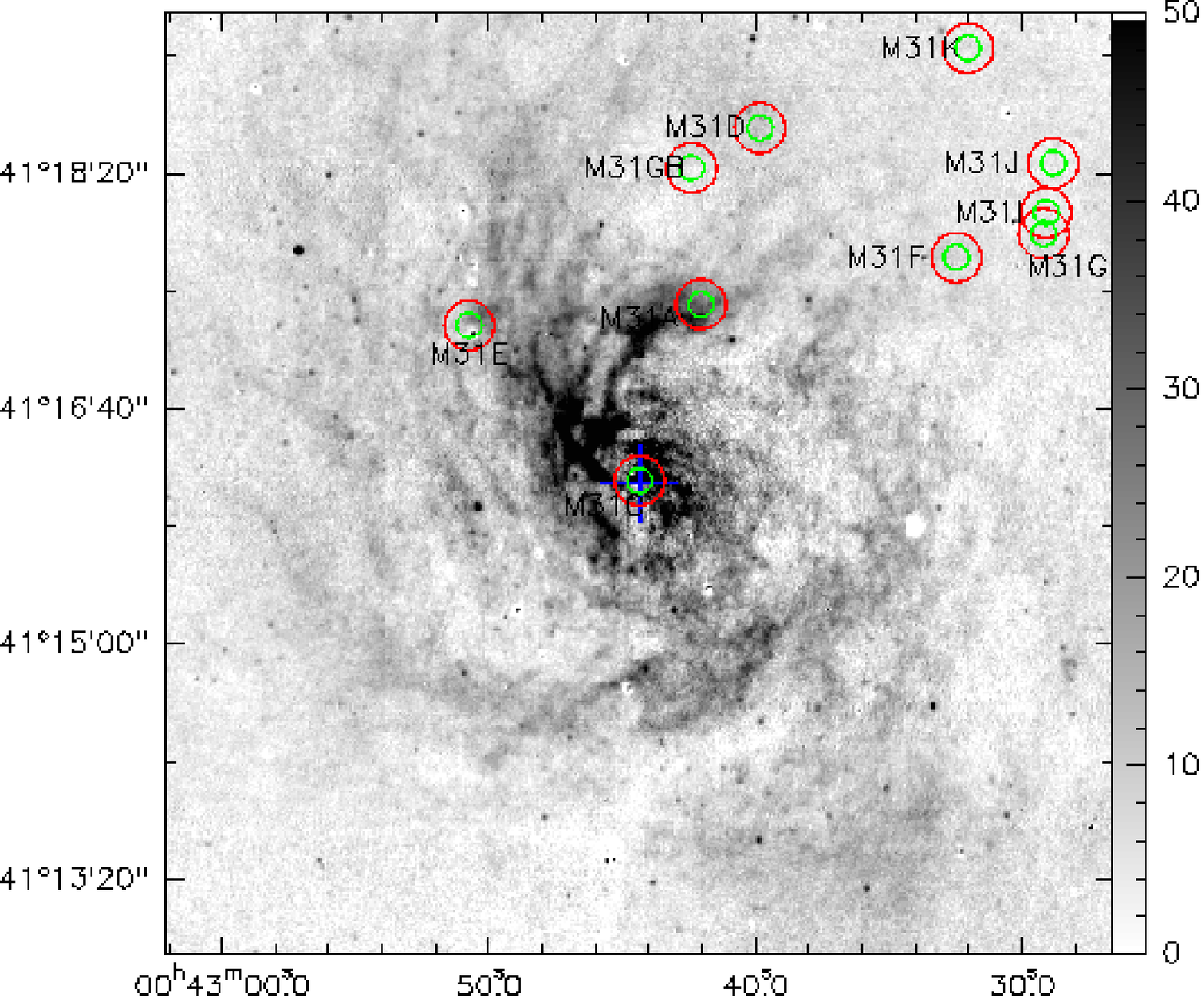}
\hspace{0.2cm}
   \includegraphics[width=6.5cm,angle=-90]{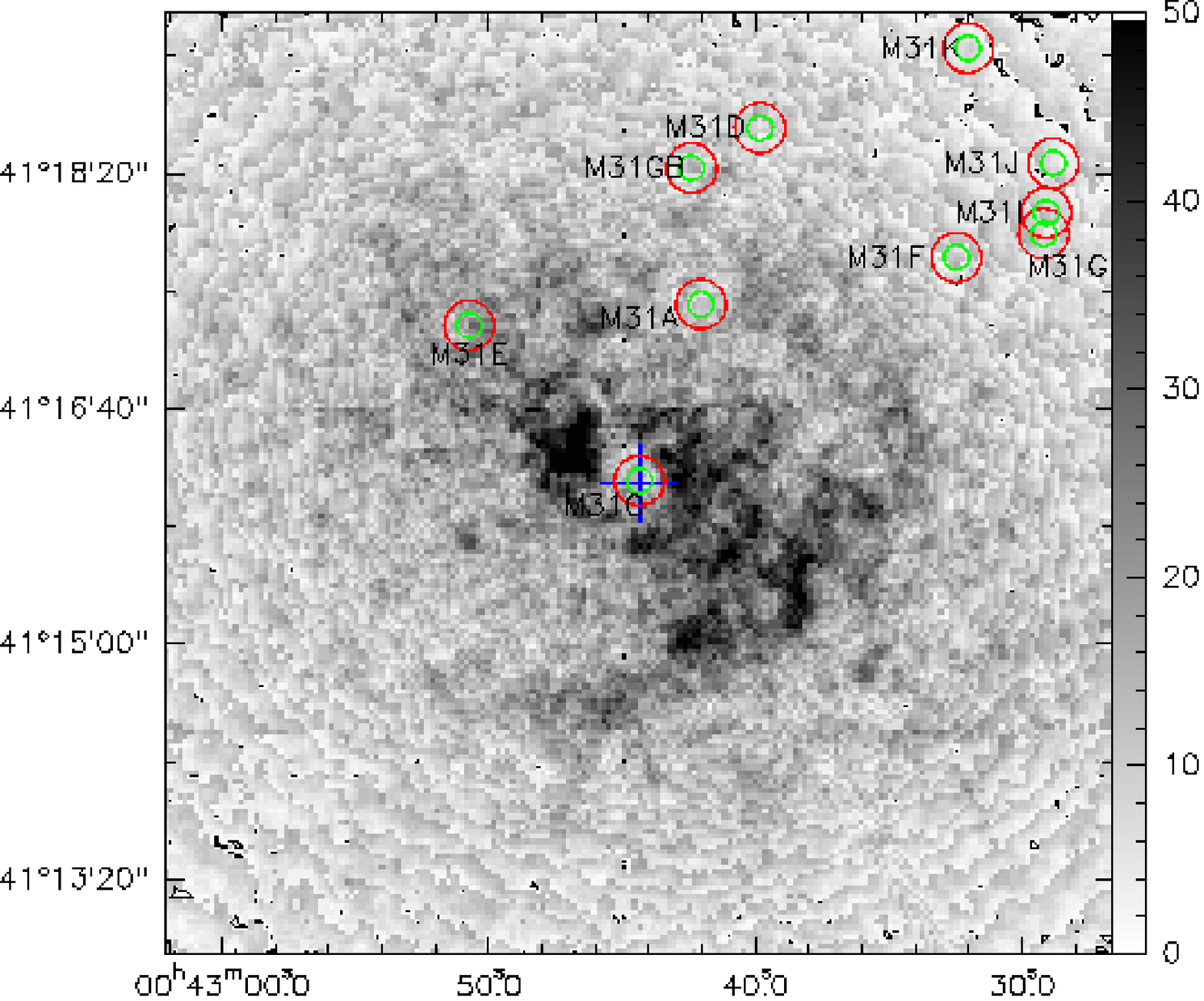}
\centering 
   \includegraphics[width=6.5cm,angle=-90]{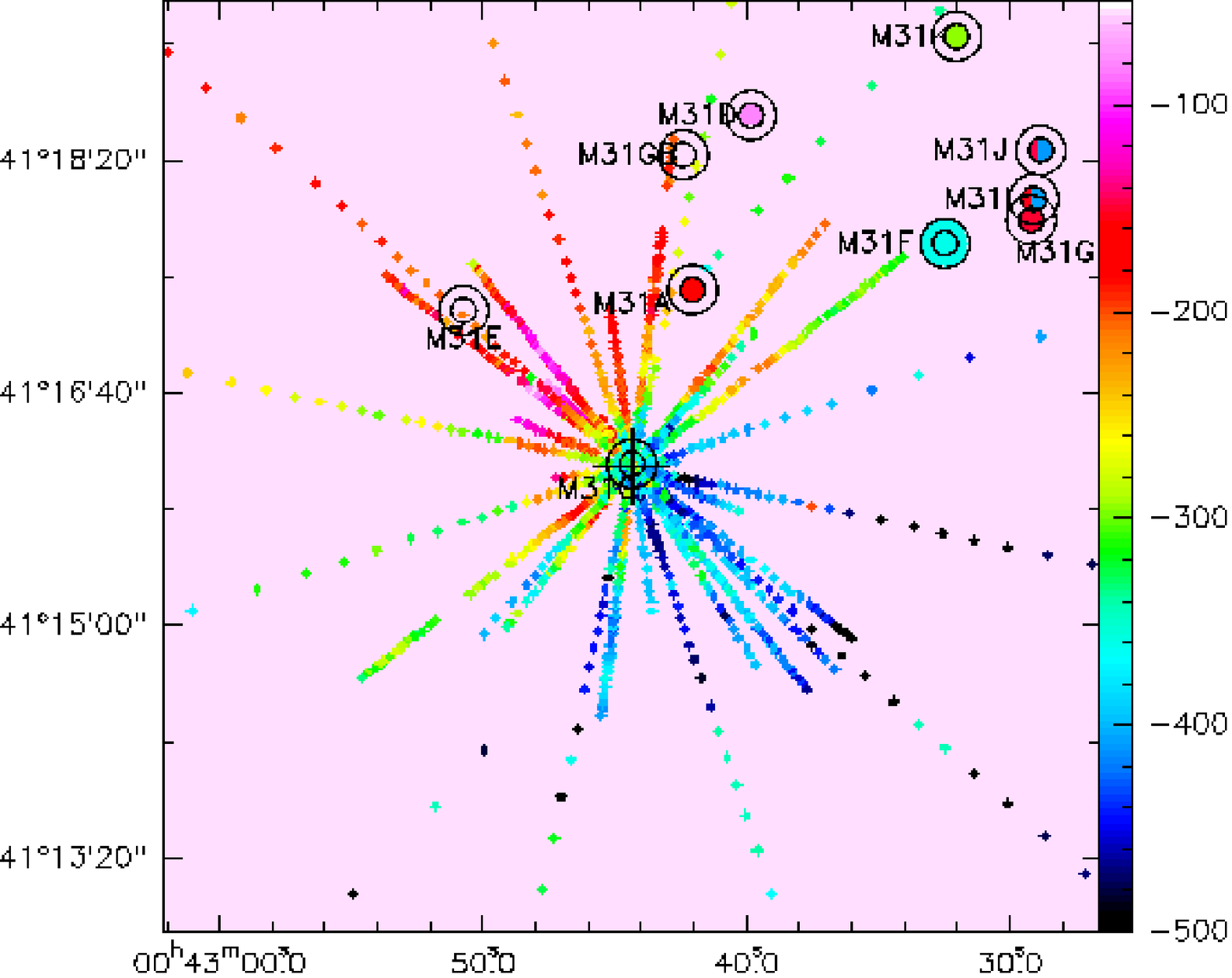}
\hspace{0.2cm}
   \includegraphics[width=6.5cm,angle=-90]{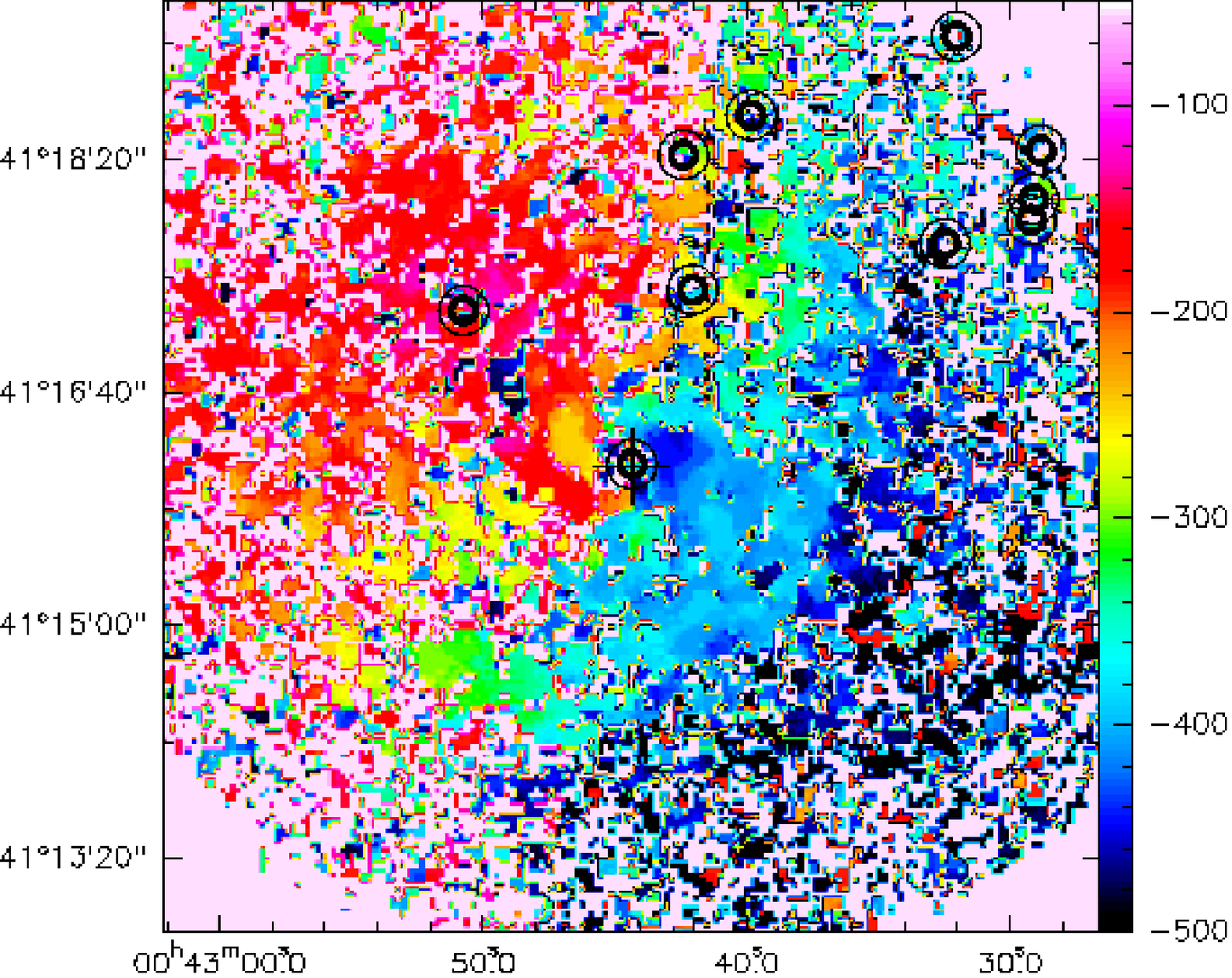}
   \caption{H$\alpha$+[NII] (left) and [NII] (right) intensities from
   \citet{Ciardullo:1988} and \citet{Boulesteix:1987} are displayed
   with an arbitrary normalisation in the top panels.  The bottom left
   panel presents the CO velocities measured in this paper,
   superimposed on the gas kinematics based on (1) H$\beta$ and
   $[$O{\sc iii}$]$ lines measured in slits by \citet{Saglia:2010},
   (2) $[$O{\sc ii}$]$ and $[$Ne{\sc iii}$]$ measured in slits by
   \citet{Ciardullo:1988} and (3) H$\alpha$ and $[$N{\sc ii}$]$
   measured in slits by \citet{Rubin:1971}. The bottom right panel
   displays the velocity field obtained with $[$N{\sc ii}$]$ $\lambda
   6584\dot{A}$ observations (cf top right panel) based on a
   Fabry-Perot device by \citet{Boulesteix:1987}. On each panel, the
   positions of our CO observations are indicated with circles
   corresponding to the beams.}  \label{fig:HaNIIvel}
\end{figure*}

\subsection{Ionised gas} 
The ionised gas in the central regions of M\,31 has been mapped by
\citet{Ciardullo:1988}, who produced the  highest S/N ratio map of
H$\alpha$+[NII] based on the compilation of a five-year nova survey
published so far. It exhibits, as displayed in the top left panel of
Figure \ref{fig:HaNIIvel}, a turbulent spiral, which appears more
face-on than does the main disc. These authors described it as lying
in a disc, warped in a way that the gas south of the nucleus is viewed
closer to face-on than the gas in the northern half of the galaxy.  As
discussed in \citet{Block:2006}, this arc feature corresponds to the
south-west part of the offset ring detected with {\em Spitzer}{-IRAC}
data. \citet{Rubin:1971} measured that in this area [NII] is three
times stronger than H$\alpha$, which supports the shock hypothesis as
also discussed by \citet{Jacoby:1985}. According to \citet{Liu:2010},
half of the kinematic temperature of hot gas in the central bulge is
accounted for by the stellar dispersion, but additional heating is
expected from type-Ia supernovae, even though the question of the iron
enrichment is not entirely clear. A head-on collision with M32
suggested by \citet{Block:2006} could account for the additional
heating and at least contribute to it.

In parallel, \citet{Boulesteix:1987} observed this same area with a
Fabry-Perot spectrograph to map the velocity field, and produced an
[NII] map, which is xdisplayed in the top right panel of Figure
\ref{fig:HaNIIvel}.  The centre has been masked for technical reasons.
Besides the lower resolution and lower S/N ratio than for the
\citet{Ciardullo:1988} map (because of  a smaller integration time), it is
instructive to observe that it exhibits different patterns compared
with the H$\alpha$+[NII] map. It is striking that the filament, which
crosses the position M31A, is not clearly detected in [NII], so it
should be mainly associated to H$\alpha$ component. The comparison of
these two maps enhances the fact that obviously forbidden lines and
Balmer lines do not sample exactly the same regions: (1) the gas does
not have the same H$\alpha$/[NII] ratio { everywhere}; (2) the
extinction additionally complicates the analysis.

\subsubsection{Velocity field of the central bulge (1.5\,kpc$\times$1.5\,kpc)} 
\begin{figure}
\centering 
   \includegraphics[width=6.5cm,angle=-90]{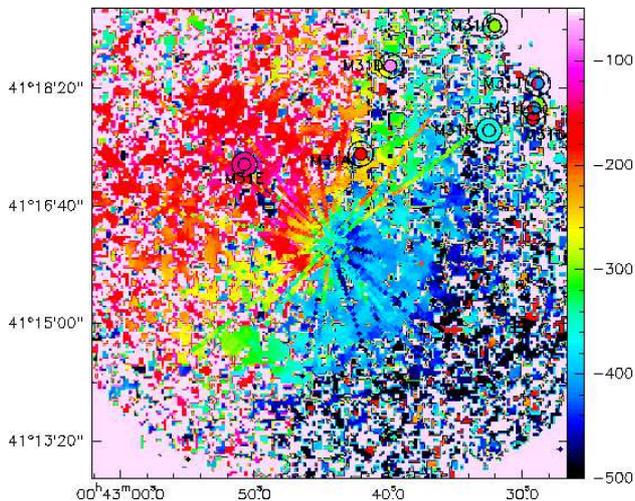}
   \caption{6.7$^{'}\times$6.7$^{'}$ ionised gas velocity field
   centred on M31. Superimposition of the various gas velocities (see
   Fig. \ref{fig:HaNIIvel} bottom left) on the \citet{Boulesteix:1987}
   Fabry-Perot-derived velocity map.}  \label{fig:XHaboul}
\end{figure}

To compare the ionised gas with our molecular detections, we tried to
reconstruct the ionised gas velocity field from data published in the
literature. Various lines have been used, but only a single velocity
has been measured for each spectrum. In the bottom right panel of
Figure \ref{fig:HaNIIvel} we display the velocity map of
\citet{Boulesteix:1987}. The best S/N ratio is of course observed
where the [NII] intensity is strongest (cf. top right panel of
Fig. \ref{fig:HaNIIvel}). It exhibits an irregular disc in rotation,
with obvious perturbations along the minor axis.  Planetary nebulae
are numerous in this field \citep[537 detected by][]{Merrett:2006} and
add noise to this velocity field. As discussed below (see Table
\ref{geometry}), we estimate here a position angle of 70$\deg$
(-20$\deg$ for the kinematic minor axis) compared to 40$\deg$ claimed
by \citet{Boulesteix:1987}. \citet{Saglia:2010} interpreted these
perturbations as counter-rotations. In the bottom left panel of Figure
\ref{fig:HaNIIvel}, we display the slit measurements we gathered from
the literature, using
Dexter\footnote{http://dc.zah.uni-heidelberg.de/sdexter} when
necessary, namely H$\beta$ and $[$O{\sc iii}$]$ from
\citet{Saglia:2010}; $[$O{\sc ii}$]$ and $[$Ne{\sc iii}$]$ from
\citet{Ciardullo:1988}; H$\alpha$ and $[$N{\sc ii}$]$ from
\citet{Rubin:1971}. We also superimpose our CO velocities, and find a very
marginal agreement.  In Figure \ref{fig:XHaboul} we superimpose the
two figures. We find a good overall agreement for the various ionised
gas measurements with complicated features in the central part and
along the minor axis, with the most striking discrepancy at position
angles of 128\,$\deg$ and 142\,$\deg$.

It is difficult to understand the discrepancies observed in the ionised
gas velocity field, as the techniques are quite different and the
\citet{Boulesteix:1987} data suffers relatively low resolution (even
though excellent given the fact that it has been obtained in 1985!).
We can consider several explanations: (1) the various lines might have
different relative ratios from one region to another  (as seen in
the top panels of Figure \ref{fig:HaNIIvel}) and might be affected
differently by extinction  (see Figure \ref{fig:super}).  (2) It
is possible that the ionised gas, like the molecular gas, exhibits
several velocity components, and it has been not explored so far, but
by \citet{Boulesteix:1987}, who mentioned line splittings greater than
30\,\kms in the central region.

In addition, the CO velocities do not follow the regular pattern. M31A
and M31D do not really match the ionised gas velocities. However, some
CO positions might have a velocity (or at least one component)
compatible with the ionised gas, namely M31E, M31F, M31K, M31I, M31J
and M31G. This suggests that only part of the molecular gas is
kinetically decoupled from the ionised gas.  

 Also, we wanted to figure out whether similar line splittings were
present in the ionised data. As discussed by \citet{Saglia:2010}, the
intrinsic velocity dispersions of the gas is smaller than 80\,\kms,
while the instrumental resolution achieved by these authors is
57\,\kms. \citet{Boulesteix:1987}, who detected line splittings larger
than 30\,\kms\, in the central area, reached a spectral resolution of
14\,\kms. It is thus challenging to detect line splittings of this
order of magnitude, but such splittings might be underlying and
explain part of the discrepancies. As we observe large ($>200$\,\kms)
line splittings in CO, we try to investigate if such double components
can be detected from the ionised gas.
\begin{table*}
	\caption[]{Line splittings detected on $[$O{\sc iii}$]$ and H$\beta$
lines from \citet{Saglia:2010}. We provide the velocities,
signal-to-noise ratio of the integrated line and $[$O{\sc iii}$]$/H$\beta$ line
ratio {($\mathbf{\log_{10}}$)} for each detected component (1 and 2). We also give for
comparison purposes the single values published by \citet{Saglia:2010}.}  
	\label{tab:linespl}
        \begin{tabular}{crr|crc|crc|c}
            \hline
            \noalign{\smallskip}
            offsets & R($\arcsec$) & PA($\deg$) &
v$_1^{[\mathrm{O{\sc iii}}]5007}$ (\kms) & S/N$_1$ & $[$O{\sc iii}$]$/H$\beta$
& v$_2^{[\mathrm{O{\sc iii}}]5007}$ (\kms) & S/N$_2$ &
$[$O{\sc iii}$]$/H$\beta$ & v$^{\mathrm{gas}}_{\mathrm{Saglia}}$ \\
            \noalign{\smallskip}
            \hline
-15.2,16.9 & -22.8& 138 &-548.6$\pm$12.0& 4.7  & 0.51$^{+0.20}_{-0.19}$ & -360.6$\pm$16.7& 5.2&-0.22$^{+0.10}_{-0.12}$&-469.1$\pm$8.4\\
-13.5,-12.2& -18.2&  48 &-462.8$\pm$6.0 & 13.1 & 0.29$^{+0.09}_{-0.08}$ & -234.0$\pm$4.0 &15.4& 0.79$^{+0.37}_{0.22}$ &-323.7$\pm$7.4\\
-12.0,-10.8& -16.2&  48 &-449.5$\pm$9.1 & 10.4 & 0.73$^{+0.20}_{-0.16}$ & -225.4$\pm$5.2 &12.2& 0.86$^{+0.32}_{0.21}$ &-314.8$\pm$5.4\\
 -9.3,-8.4 & -12.5&  48 &-485.8$\pm$9.9 &  4.9 & 1.17$^{+\inf}_{-0.71}$ & -375.0$\pm$6.7 & 3.9&-0.23$^{+0.20}_{0.21}$ &-418.5$\pm$10.7\\
 -8.1,-7.3 & -10.9&  48 &-454.3$\pm$14.4&  3.3 & 0.15$^{+0.23}_{-0.25}$ & -272.5$\pm$13.3& 5.3& 0.41$^{+0.22}_{0.20}$ &-314.4$\pm$9.2\\
 -1.9,-5.9 & -6.2 &  18 &-319.9$\pm$20.0& 3.3  & 0.14$^{+0.34}_{-0.31}$ & -144.8$\pm$15.9& 3.6 & 0.19$^{+0.32}_{0.28}$&-342.7\\
 -0.2,0.9  & -0.9 & 168 &-281.3$\pm$8.9 & 8.7  & 0.78$^{+1.05}_{-0.33}$ & -85.9$\pm$16.2 & 3.2 & 0.00$^{+0.72}_{0.41}$&-241.8$\pm$10.0\\
  0.1,-0.7 &  0.7 & 168 &-319.9$\pm$19.8& 3.3  &-0.20$^{+0.14}_{-0.18}$ & -144.8$\pm$16.1& 3.7 &-0.04$^{+0.20}_{0.22}$&-221.7$\pm$11.6\\
  0.7,0.6  &  1.0 &  48 &-332.6$\pm$11.0&  6.8 & 0.60$^{+0.22}_{-0.18}$ &  -64.9$\pm$7.5 & 7.5& 1.04$^{+\inf}_{-0.40}$&-313.6$\pm$7.8\\
  1.1,1.0  &  1.5 &  48 &-322.3$\pm$13.0&  4.9 &-0.06$^{+0.17}_{-0.17}$ &  -56.0$\pm$3.5 &18.2& 0.92$^{+0.32}_{0.20}$ &-330.6$\pm$7.8\\
  1.2,0.3  &  1.2 &  78 &-321.5$\pm$13.2&  5.2 & 0.20$^{+0.26}_{-0.22}$ &  -78.7$\pm$4.8 &12.7& 0.57$^{+0.22}_{0.17}$ &-327.8$\pm$11.9\\
  8.8,-2.9 &  9.3 & 108 &-598.0$\pm$4.7 & 14.1 &0.92$^{+0.37}_{-0.22}$ & -319.2$\pm$14.0& 8.1 & 0.18$^{+0.12}_{-0.12}$&-343.6$\pm$13.9\\
 11.2,-3.6 & 11.7 & 108 &-593.7$\pm$5.5 & 8.4  &1.17$^{+0.56}_{-0.28}$ & -373.3$\pm$29.2& 3.7 &-0.18$^{+0.16}_{-0.19}$&-376.8$\pm$18.1\\
  5.3,16.4 & 17.2 &  18 &-575.1$\pm$3.8 & 6.4  &-0.16$^{+0.17}_{-0.16}$ & -444.8$\pm$35.2& 4.5 &0.28$^{+0.17}_{-0.18}$&-291.4$\pm$6.6\\
            \noalign{\smallskip}
            \noalign{\smallskip}
            \hline
         \end{tabular}
\end{table*}
\begin{figure}
\centering 
   \includegraphics[width=6.5cm,angle=-90]{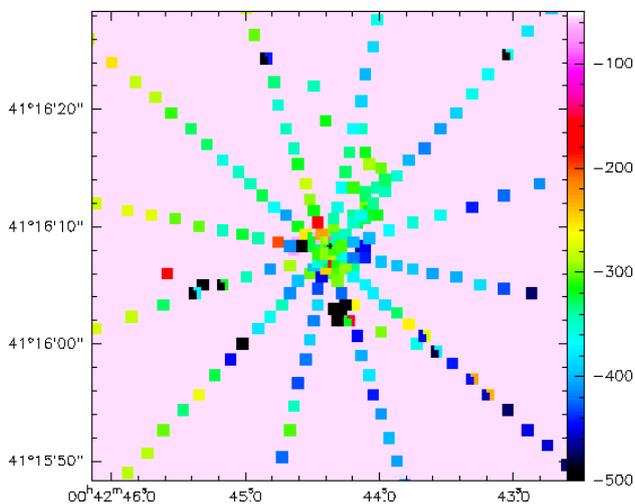}
   \caption{40$^{''}\times$40$^{''}$ ionised gas velocity field
   centred on M31. Line splittings detected in $[$O{\sc
   iii}$]$\,50007$\dot{A}$  are superimposed on the velocity field
   measured on ionised gas by \citet{Saglia:2010} and by
   \citet{DelBurgo:2000}. This velocity field is quite unexpected and
   do not display any clear rotation pattern. } \label{fig:HLS}
\end{figure}

Optical spectra of M\,31 are heavily dominated by the stellar
continuum of its bulge. Therefore, in order to study emission lines in
that spectral region we approximated the calibrated sky subtracted
spectra of M\,31 kindly provided by R.~Saglia by stellar population
models using the {\sc NBursts} full spectral fitting technique
\citep{CPSA07,CPSK07}. We excluded narrow regions around H$\beta$,
$[$O{\sc iii}$]$, and $[$N{\sc i}$]$ lines from the fitting. Then we
subtracted the best-fitting stellar population models from the
original spectra and analysed the fitting residuals. This allowed us
to reliably measure parameters of these rather faint emission lines
which are barely visible in the original data. We then inspect
visually the spectra to determine the positions where high S/N line
splittings is present. We then fit a two Gaussian functions with the
CLASS package on the $[$O{\sc iii}$]$5007$\dot{A}$ line. Template mismatch and
the relative low S/N affect the measurement of the H$\beta$, which is
weaker and not always detected. For none of our measurements (but
one), the [NI] line is detected. Our detection are summarised in Table
\ref{tab:linespl} and Figure \ref{fig:HLS}. We identify 14 positions
with a double component detected in $[$O{\sc iii}$]$ and also in
H$\beta$, within 23\arcsec\, ($\sim$86\,pc) from the centre, but we do not
detect any line splittings close to our CO observations or at a
similar radial distance.

\subsubsection{ Velocity field of the circumnuclear region
(150\,pc$\times$150\,pc)}  
\label{sssect:linesplit}
 Figure \ref{fig:HLS} displays the velocity field measured in
ionised gas of the circumnuclear region. This includes slit data from
\citet{Saglia:2010} as well as two-dimensional spectroscopy of
\citet{DelBurgo:2000} for the measurements of single velocity
components, and our line splitting measurements based on 	 
\citet{Saglia:2010} data. It is striking that
unlike Figure \ref{fig:XHaboul} this velocity field does not exhibit
any clear rotation pattern in the central field. While the centre
presents a spot at the systemic velocity, the whole area has
velocities smaller than the systemic velocity and is approaching the
observer. This coherent flow of ionised gas is decoupled from the
stellar kinematics
\citep[e.g.][]{Bender:2005,DelBurgo:2000,Saglia:2010}, and resembles
an ionised gas outflow. This is confirmed by the fact that the two
independent data sets are in good agreement. { We should stress
that the \citet{Saglia:2010} data are characterised by an instrumental
resolution of 57\,km\,s$^{-1}$, and their systemic velocity is shifted
by $+23$\,km\,s$^{-1}$ with respect to the other studies. However, we
checked that this cannot account for the systematic blue shift
observed with respect to the centre.} The mechanisms that rule the
feed-back from the AGN are still largely unknown and in the case of
M31, we are only tracing a relic activity of its known black
hole. Considering that M\,31's black hole is not active, two main
mechanisms can be considered to explain this outflow. (1) It could be
directly associated to the past stellar activity detected in the
central regions: a 200Myr old stellar disc with a 2400$\msol$ mass has
been detected by \citet{Bender:2005} within a fraction of arcsec from
the centre. Inside a 2\arcsec disc,
\citet{Saglia:2010} has also detected a stellar component younger than
600\,Myr corresponding to a mass smaller than 2$\times 10^{6}$\msol:
it could be compatible with a 100\,Myr component with a mass of
$10^{6}$\msol. However, {we lack of constraints to conclude for} a
star formation origin {with a typical} dynamic time $\tau_{dyn} = $
size $/$ velocity of this outflow. {The typical velocity of an outflow
\citep[e.g.][]{Rupke:2005} is about 400\,km\,s$^{-1}$, a
$\tau_{dyn}=100$\,Myr old starburst would then correspond to a size of
the emitting area of 40\,kpc.  It might correspond to the X-ray
outflow detected by \citet{Bogdan:2008} (see Figure \ref{fig:XHa},
left panel).} (2) As discussed by various authors
\citep[e.g.][]{David:2006,Ho:2009}, the ionised gas in galaxy bulges
can be accounted for by mass loss of evolved stars and SNIa could play
a key role. The prompt SNIa component (which represents 50\% of the
SNIa population) discussed by \citet{Mannucci:2006} could hence be
indirectly linked to the 100-200Myr star formation episodes quoted
above { and contribute to this outflow}.

As discussed in the following, a possible collision with M32 can
explain the ring structures of M31 where the CO has been detected, and
such an event could have triggered a star formation activity in the
central part and support the SNIa explanation for the outflow.  In
addition, this inner outflow could be connected with the outflow
detected in X-ray by \citep{Bogdan:2008} on a larger scale and
discussed in Sect. \ref{ssect:other}.  
\begin{figure*}
\centering 
   \includegraphics[width=6.5cm,angle=-90]{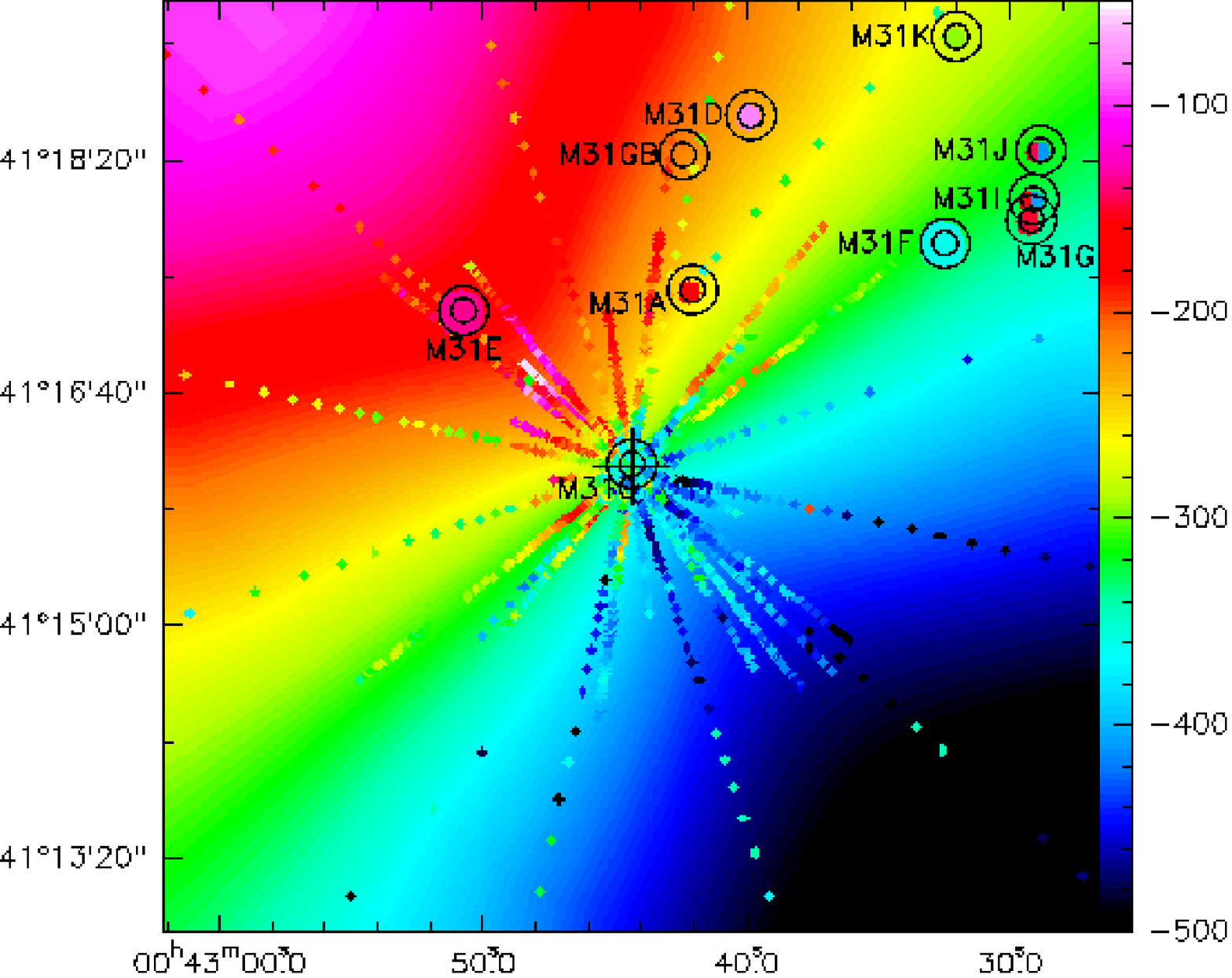}
   \includegraphics[width=6.5cm,angle=-90]{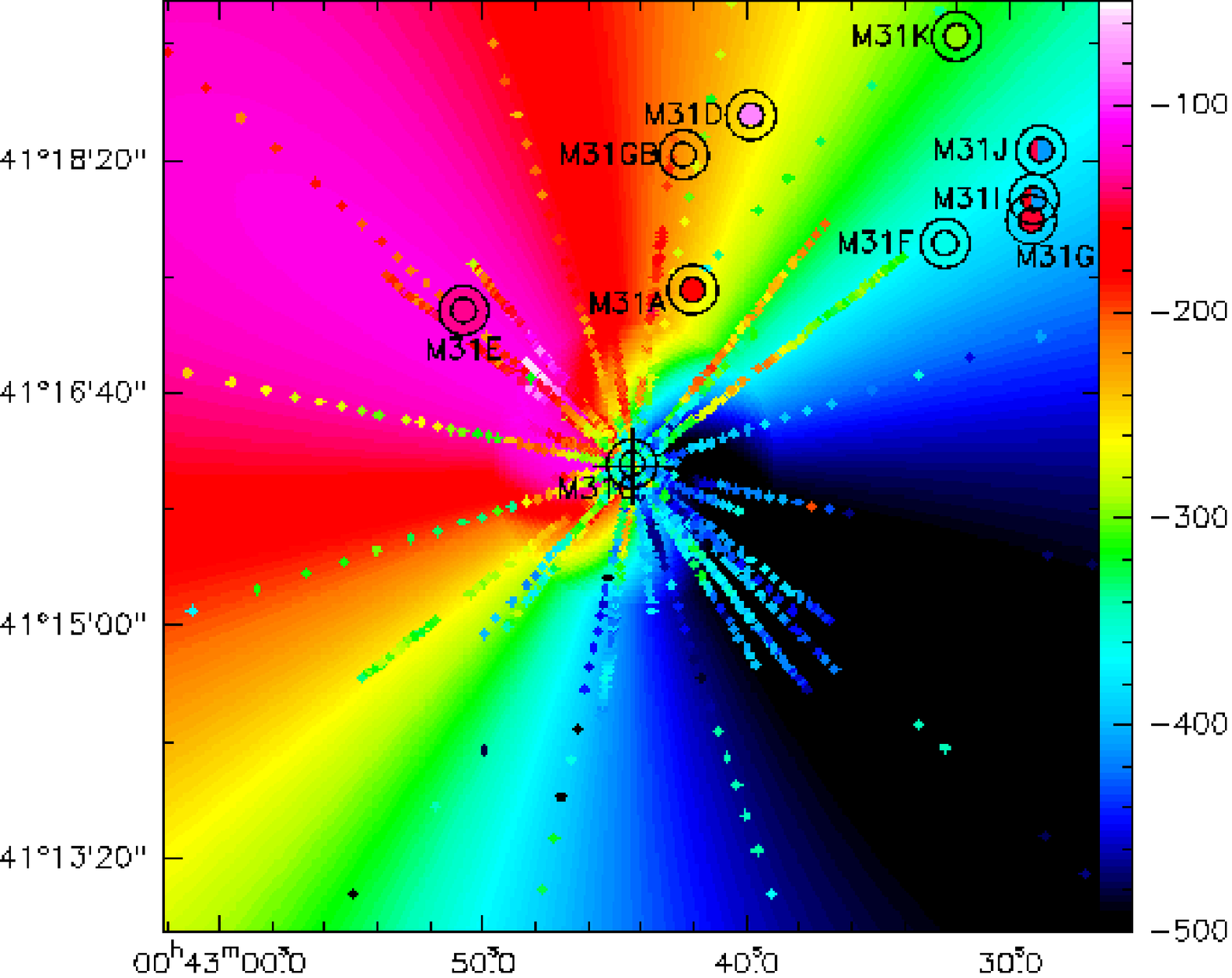}
   \caption{Superposition of the various gas velocities (see
   Fig. \ref{fig:HaNIIvel} bottom left) (left:) on a simple model of a
   galactic disc with an inclination of 45$\deg$, a position angle of
   {40$\deg$} and a systemic velocity of -310\,\kms; (right:)
   similar disc with variable position angles} \label{fig:XHamod}
\end{figure*}

Several (8/14) of the positions with line splittings have a component
with a large $[$O{\sc iii}$]$/H$\beta$ ratio. Such a large ionisation is
compatible with planetary nebulae. \citet{DelBurgo:2000} observed
several high-ionisation ``clouds'' in this area. These authors
discussed that the intensity of 3 of these sources is much larger than
those of planetary nebulae detected in M31 by \citet{Ciardullo:2002}.
However, in this central region, it might be possible that these sources
are multiple. Interestingly, one of the sources (D) detected by
\citet{DelBurgo:2000} exhibits a line splitting
(-270,-527)\,km\,s$^{-1}$ with an amplitude comparable with ours but
very close to the centre, even though none of the velocities really
match. Also, the strongest component we detect in the inner arcsec
region matches in first approximation with the source A of
\citet{DelBurgo:2000}.
\begin{figure}
\centering 
   \includegraphics[width=6.5cm,angle=-90]{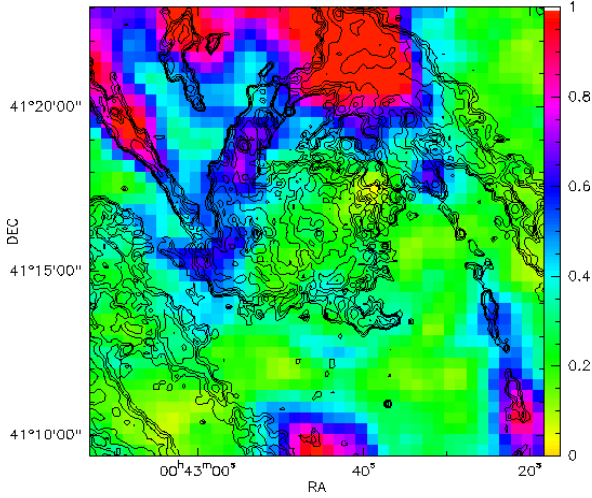}
   \caption{Image of the HI emission in the very centre, avoiding
   $\pm$100km/s around the systemic velocity of -310km/s.  This cut
   has been made to subtract the large-scale HI emission seen in
   projection superposed to the centre, expected near the systemic
   velocity. The contours are the dust emission from {\em
   Spitzer}-IRAC { (8$\mu$m image where a scaled version of the 3.6
   $\mu$m image has been subtracted to remove the stellar photospheres
   emission)} \protect\citep{Block:2006}. The HI cube is from Braun et
   al (2009).  }  \label{fig:HI-dust}
\end{figure}
\begin{figure}
\centering 
   \includegraphics[width=6.5cm,angle=-90]{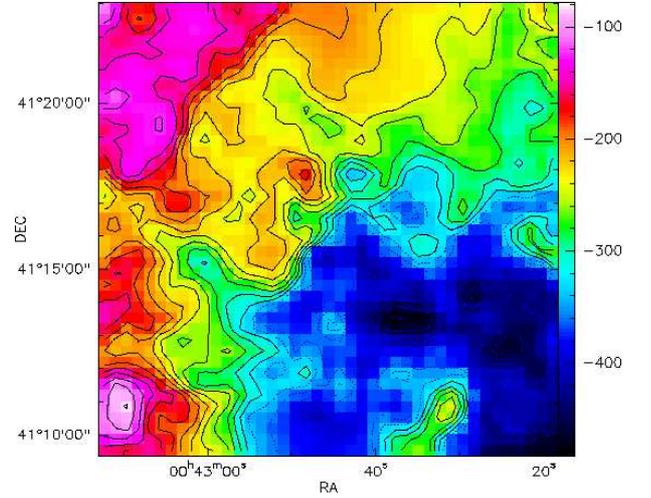}
   \caption{Velocity field of the HI emission, corresponding to 
{Figure \ref{fig:HI-dust},} avoiding large-scale HI emission.
  }  \label{fig:velo-both}
\end{figure}

{A comparison with SAURON data (M. Sarzi, private communication)
taken in this region of M31 has shown that our multiple lines
detections are compatible with planetary nebulae. The slit most
probably biases the line ratio determination, as different parts of
the PSF are sampled.}

\subsubsection{Modelling of a tilted inclined disc}
We create a velocity field map with the CCDVEL task within the NEMO
software package \citep{Teuben:1995}. Following
\citet{Ciardullo:1988}, we consider a tilted ring model with an
inclination of 45$\deg$, a position angle of {40$\deg$ 
\citep{Boulesteix:1987}} and a systemic velocity of -310\,\kms. In
order to understand the behaviour of the gas in the central part of
M31, we superimpose the velocities measured in this paper (ionised gas
and CO gas) on this modelling in Figure \ref{fig:XHamod} {(left
panel)}. Even though there is a velocity field, as shown with ionised
gas by \citet{Boulesteix:1987} and displayed in the bottom right panel
of Figure \ref{fig:XHaboul}, it is not regular and does not exhibit a
well-defined zero-velocity curve. In addition, the velocities of the
CO measurements do not fit with such a regular speed pattern. Varying
the velocity angle in the centre, {as displayed in the right panel
of Figure \ref{fig:XHamod},} mimics the velocity pattern in the
inner region ($<10$\,arcsec) but does not explain the minor axis
configuration.

{\subsection{HI emission} The detailed maps from \citet{Braun:2009}
have revealed clearly the HI deficiency in the central parts of M31.
There is, however, some HI emission in the central kiloparsec, but
most of it comes near the systemic velocity, and is likely to be
projected emission from the external tilted gas orbits, which are
warped to an almost edge-on inclination \citep[e.g.][]{Corbelli:2010}.
To subtract this projected large-radii emission, and better see the
residual coming from the actual centre, with a large velocity
gradient, we have summed all channels avoiding $\pm$100km/s around the
systemic velocity of -310km/s. A weak signal can then be
distinguished, with a morphology of an incomplete ring {(Figure
\ref{fig:HI-dust})}, with emission corresponding to the east side of
the inner ring delineated by the dust emission revealed in the {
8$\mu$m} {\em Spitzer}{-IRAC} map.  Figure \ref{fig:HI-dust} shows the dust
contours superposed on the residual high-velocity central HI emission.
In this picture, we can see that the main HI residual component still
follows the large-scale (NE-SW) arms seen in projection, and
coinciding with the dust features, in the direction of the major axis
of M31. However, there remains a weak component perpendicular (NW-SE)
to it.  The main concentration of this residual HI emission is well
aligned with the north-east part of the dust ring and it could
correspond to the weak component seen by \citet{Brinks:1983} on the
minor axis position velocity diagram (his figure 1b).  However, the
remaining parts of the ring is devoid of HI emission; the atomic gas
must have been transformed into the molecular phase in the dense parts
of the ring.  { As discussed in Appendix \ref{sect:append}, Figure
\ref{fig:HIvel}} reveals that most of the CO strong emission at high
velocity in the centre has no HI counterpart.

The HI component associated to the dust ring is, however, too weak to
be seen in the velocity map, as shown from Figure
\ref{fig:velo-both}. The velocity field in the HI selected with high
speed with respect to the systemic velocity, is still compatible to
the ``normal'' {main} disc, and similar to the velocity field
found with the ionised gas in H$\alpha$, although with {possible}
perturbations in the vicinity of the ring. {It is difficult to
determine an accurate position angle of the HI isovelocity map (Figure
\ref{fig:velo-both}) to discriminate between the
nuclear disc and the main disc, but the intensity map (Figure
\ref{fig:HI-dust}) does not really favour a nuclear disc component.}

\subsection{Other wavelengths information}
\label{ssect:other}
\begin{figure*}
\centering 
   \includegraphics[width=6.5cm,angle=-90]{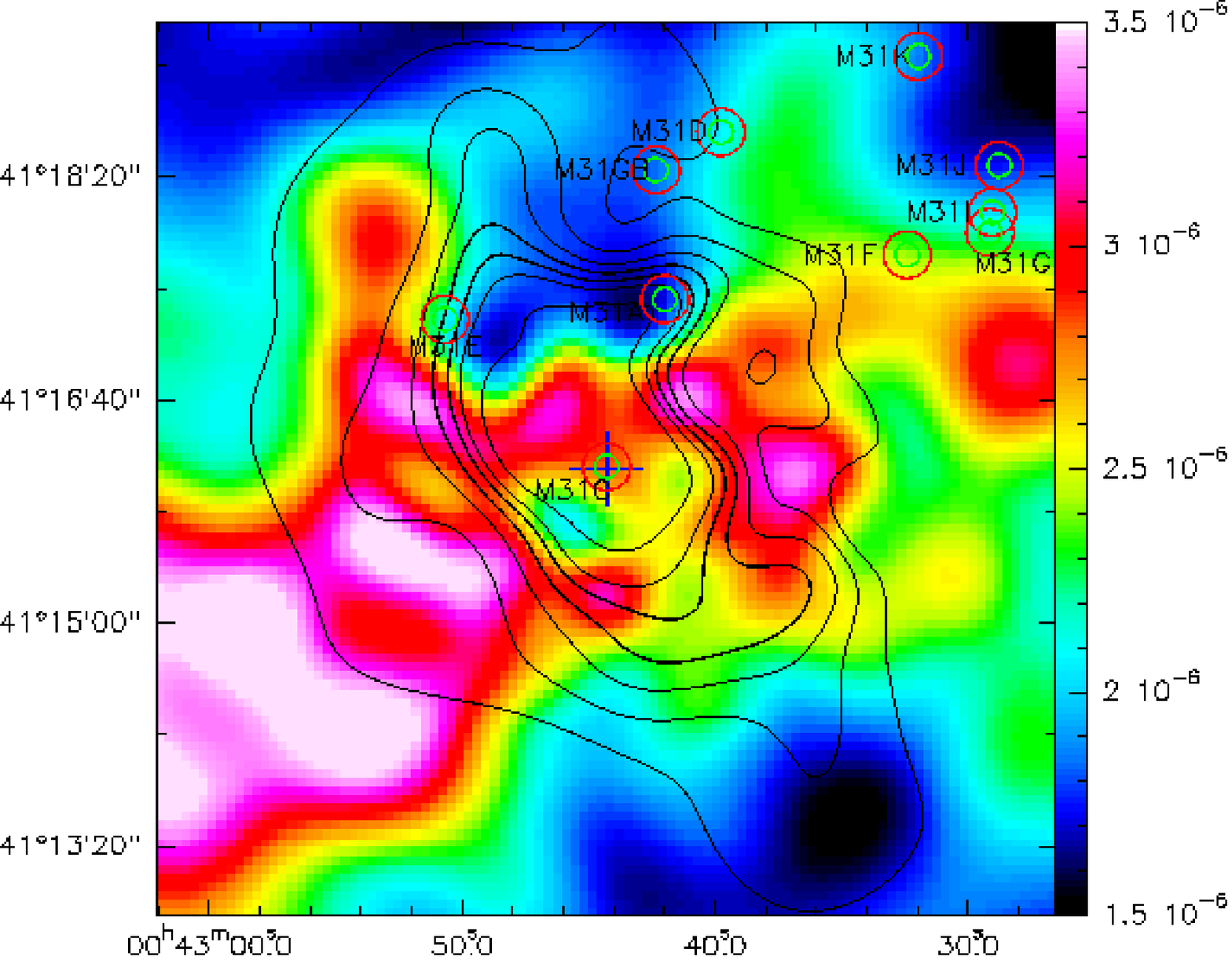}
   \includegraphics[width=6.5cm,angle=-90]{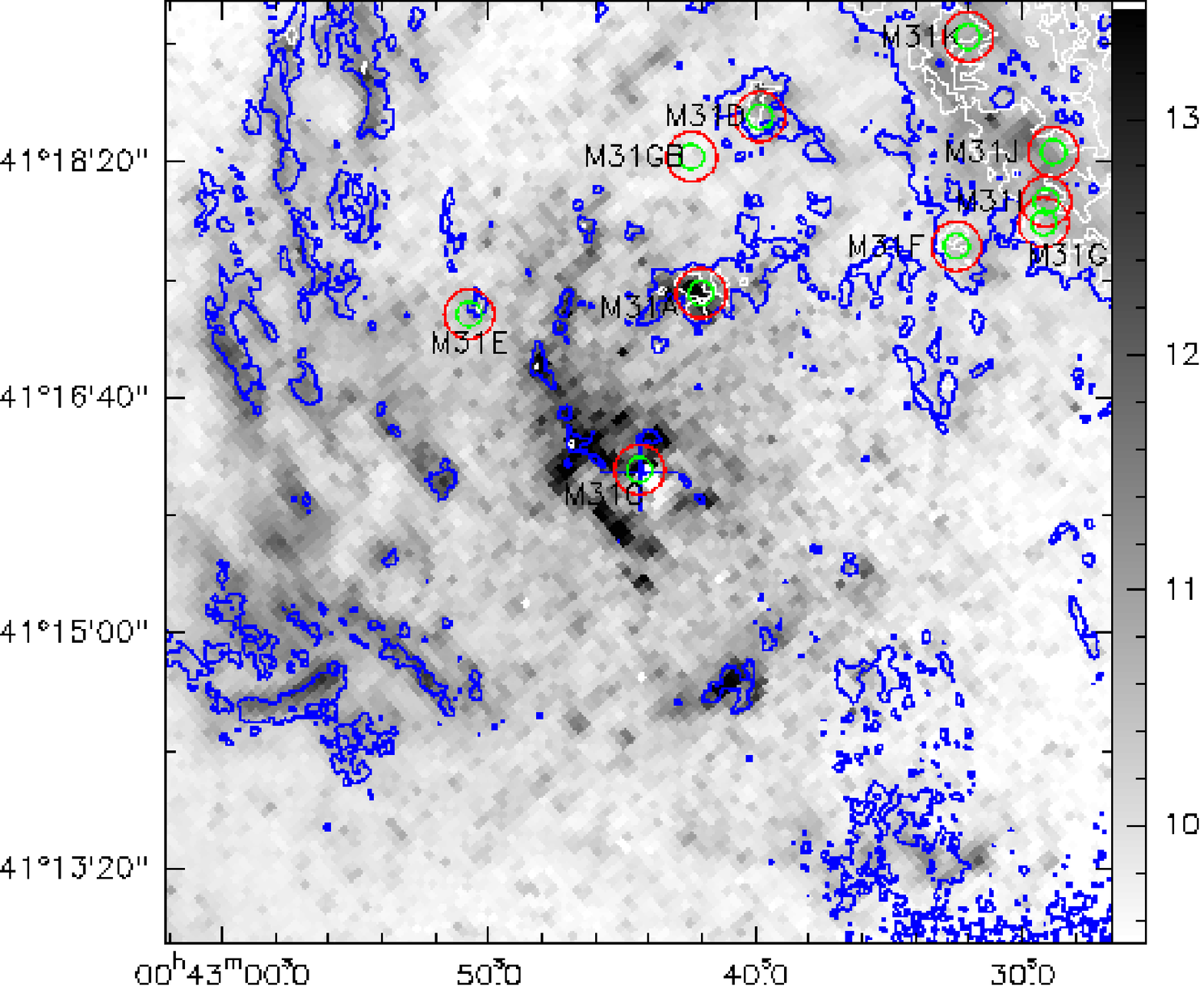}
   \caption{Left: Superposition of the H$\alpha$+$[$N{\sc ii}$]$ map
   (upper left panel of figure \ref{fig:HaNIIvel}) of
   \citet{Ciardullo:1988} on the Chandra soft X-ray emission from
   \citet[see their Figure 7]{Bogdan:2008}. The H$\alpha$+$[$N{\sc
   ii}$]$ map has been smoothed with the same smoothing lengths as
   those used for the Chandra map (A. Bogdan, private
   communication). Right: Superposition of the $A_B$ extinction on the
   8$\mu$m {\em Spitzer}{-IRAC} map after subtraction of a scaled version
   of the 3.6 $\mu$m image { \citep{Block:2006}}. The $A_B$
   contours are fixed to 0.05 (blue) and 0.15 (white).}
   \label{fig:XHa}
\end{figure*}
While the centre of M31 hosts a massive black hole with a mass of
$0.7-1.4 \times 10^8 M_\odot$ \citep{Bacon:2001, Bender:2005}, it is
one of the most underluminous supermassive black hole
\citep{Garcia:2010}. The A-star cluster, detected in the { third}
component { (P3) of M\,31's nucleus by \citet{Bender:2005}}, can be
associated to a recent star formation episode, { typically} a
single burst, which occurred 200\,Myr ago, with a total mass in the
range 10$^4$ -- 10$^6 M_\odot$, corresponding to an accretion rate of
10$^{-4}$ -- 10$^{-2} M_\odot$\,yr$^{-1}$. It might be triggered by
the possible frontal collision with M\,32 \citep{Block:2006}. The
detection of an ionised gas outflow in X-rays along the minor axis of
the galaxy by
\citet{Bogdan:2008}, perpendicular to the main disc, could be linked
to this recent star formation activity and to the possible outflow
detected in the optical (Figure \ref{fig:HLS}). However, this is still
controversial as such a burst does not have enough energy to power the
galactic wind required \citep{Bogdan:2008}. In the left panel of
Figure \ref{fig:XHa}, H$\alpha$+$[$N{\sc ii}$]$ contours are
superimposed (with the same resolution) on the Chandra soft X-ray
emission map from \citet{Bogdan:2008}. The relative intensity of the
outflow on both sides is compatible with the intensity of A$_B$
extinction: the NW side is more extinguished than the SE side. This is
compatible with the modelling described in the next section.  Last, the
right panel of Figure \ref{fig:XHa} displays superimposition of the
contours of the A$_B$ extinction map on the PAH-dust emission at
8\,$\mu$m detected by {\em Spitzer}. The extinction features match
exactly the ring detected at 8\,$\mu$m.

\section{Interpretation}
\label{sect:inter}
{The molecular, atomic and ionised gas exhibit different radial
distributions in disc galaxies as first discussed by
\citet{Kennicutt:1989} \citep[see also][for a recent
review]{Bigiel:2008}. The HI gas is known to extend at much larger
radius. M31 has HI gas extending at least up to 40\,kpc
\citep{Corbelli:2010,Chemin:2009} with high velocity clouds up to
50\,kpc \citep{Westmeier:2008}. As discussed below, the projected warp dominates
the HI emission in the central part (see also
\ref{sect:append}). The CO and ionised gas are usually more
concentrated. In M31, while CO is known to be depleted in its centre
\citep{Nieten:2006}, the ionised gas is detected in the inner
parts \citep{Ciardullo:1988,Boulesteix:1987} and is strongest there
\citep{Devereux:1994}.  }

{While the atomic and ionised gas exhibit the presence a perturbed
disc, the molecular gas detected in CO displays unexpected kinematic
signatures, with significant line splits close to the minor axis and a
very weak when present signal close to the systemic velocity.  Several
scenarios could be invoked to explain the existence of two well
separated high S/N velocity components ($\Delta$V = 260\,\kms
on a scale of 40\,pc) traced by the molecular component (CO emission)
in the centre of M31. One component is in the sense of the expected
rotation, the other component is counter-rotating. The widths of these
two components are also very different, as displayed in Figures
\ref{fig:M31all} and \ref{fig:superimp}. }

 The ionised gas also exhibits unexpected kinematic features with a
face-on outflow in the circumnuclear region (detected in the optical),
extending (in Xray) to the 1.5\,kpc $\times$ 1.5\,kpc region studied
in this paper.

 In the following, we explore  several possible scenarios to
explain the observations.  In Sect. \ref{ssect:warploop}, we
discuss the expected effects of a large scale warp.  In
Sect. \ref{ssect:bar}, we summarise the expected signatures of a bar
and demonstrate that they fail to match the CO observations.  In
Sect. \ref{ssect:ring}, we remind the scenario of a head-on collision
with M\,32 as proposed by \citet{Block:2006} and show with a simple
modelling that it can account for the various observables. 
\label{ssect:scenario}

\subsection{Scenario 1: large scale structures warp and tidal streams}
\label{ssect:warploop}
 We discuss the arguments that the second velocity component
detected in CO could be due to the superimposition of gas at large
scale onto the central regions. These possible configurations are not
related to the outflows detected there in the ionised gas.

\subsubsection{Large scale warp}
{We could think that the second component, the counter-rotating one,
is just observed in projection, and comes from the external warp
observed in the outer parts of the M31 disc, with a different
inclination and position angle than the normal M31 disc. The warp is
visually conspicuous, and remarkable in HI
\citep[][]{Corbelli:2010}. Farther in the north-east disc (beyond
10\,kpc), \citet{Casoli:1988} already noted that CO emission could
come from the main disc and the warped component, both on the same
line of sight. In that case, the CO emission ($T_A^* \sim 40mK$) was
coming from material distant by more than 16\,kpc from the centre.

This scenario is, however, impossible, because (1) the expected
molecular gas has negligible emission at the large radii where this
warped component should be \citep[e.g.][]{Neininger:1998}; (2) the
corresponding projected velocity on the line of sight close to the
centre is expected around the systemic velocity. It is likely that the
gas at long distance (16-30\,kpc) is on nearly circular orbits and has
no radial velocity when projected to the centre. Accordingly, the
corresponding warped material was always found at systemic
velocity. Whatever inclination it has, the gradient in velocity across
a small region of 40pc should be negligible. Ad hoc hypotheses with HI
infall/outflow with important radial velocity are then required, which
are incompatible with the HI observations (e.g. Corbelli et al
2010). We therefore think that, unless the gas is completely out of
equilibrium
\citep[which is not really supported by the observations of][]{Corbelli:2010,Casoli:1988}, the peculiar component cannot come
from the outer warp. However, we have some tentative detections close
to the systemic velocity (M31K, M31I, M31G, M31J), which could be
accounted for by the external warp. Some positions (e.g. M31G) suffer
from signal subtraction caused by the off positions (wobbler mode of
observation), and we cannot exclude that the real signal could be a
bit larger.}

\subsubsection{Large-scale tidal streams}
\label{ssect:loop}
{The velocity range measured in CO is comparable with the
measurements performed by \citet{Ibata:2004,Chapman:2008} for stellar
streams and for extra-planar gas and high-velocity clouds detected in
HI outside the disc
\citep{Braun:2004,Westmeier:2008}. \citet{Casoli:1988} have detected CO
up to 16\,kpc.  \citet{Braun:2004} detected a faint bridge of HI
emission which appears to join the systemic velocities of M\,31 with
that of M\,33 and continues beyond M\,31 to the north-west. Davies's
cloud \citep{Westmeier:2008} lies in the north-west part of M31's disc
and could belong to this bridge. Similarly to the stellar streams
observed in the south-east part of M31's disc \citep{Ibata:2007}, the
HI gas exhibits large design loop-like features \citep[e.g. the
Magellanic stream,][]{Kalberla:2006}. The second component detected in
the molecular gas we have detected could be associated to the M31-M33
bridge or any gaseous-loop relics. It would then correspond to some
accretion of gas onto the disc, generating an inner polar ring and
leading to compression and excitation of the CO.  This configuration
can be compared to the inner polar gaseous disc discussed by
\citet{Sil'chenko:2011} in NGC\,7217, where the disc is face-on.

While HI studies mention contamination by the Galaxy
\citep{Westmeier:2008}, our CO detections are related to extinction
patterns that are clearly associated to M\,31. The CO detected for M31D has a
mean velocity of -74\,\kms is typically below the limit for Galactic
contamination used for HI. However, the typical velocity dispersion of HVC
in the Galaxy is 8.5\,\kms, while the dispersion measured for M31D is
24\,\kms. Furthermore, it exhibits a higher than expected
I$_{CO}$/E(B-V) ratio, rather suggesting that it lies behind the bulge
and not in front of it.  Finally, no CO from the Milky Way has been
detected by \citet{Dame:2001} in the vicinity of M31.

Last, we cannot exclude that some non-circular velocities detected in
CO result from relics of M31's formation: scattered debris or other
extra-planar gas.
}

\subsection{Scenario 2: a possible bar}
\label{ssect:bar}
Another a priori plausible scenario could be that the second
velocity component is associated to a possible bar.  Bars are frequent
in spiral galaxies \citep[60 to 75\% are barred according to the bar
strength; e.g.][]{Verley:2007}, and it is likely that M31 is
barred { or has been barred, because of its triaxial boxy bulge 
\citep{Beaton:2007}. However, up to now, the bar is not visible in the gas
in either the morphological or kinematical data.}

\subsubsection{Previous bar propositions in M31}
\label{ssect:bar1}
There have been several propositions of bar models for M31.  However,
they are all contradictory (orientation of the bar along the minor
axis or the major axis), and there is no coherent model that fits the
observations. \citet{Stark:1994} assume the existence of a Ferrers bar
potential in the centre to explain some non-circular motions there,
but there was no complete velocity field, and they do not try to fit
the observed stellar distribution with a bar.  Their main point is to
propose that the morphology of the gas in the centre of M31 looks
round, { in spite of the bar, because of the high (77$^{\circ}$)
inclination of M31: the bar would be aligned along the minor axis. The
{\it x1} orbits, elongated along the bar, i.e.  the minor axis, are
then projected as almost round, because their axis ratio could be as
high as 4.44= 1/cos(77$^{\circ}$)}. The authors propose {\it x1}
orbits between 0.7 and 3\,kpc in radius, and perpendicular {\it x2}
orbits inside 0.7\,kpc. The gas would be shocked at the transition
between {\it x1} and {\it x2} orbits, in a ring of 0.7\,kpc radius.
However, as is demonstrated below {in Sec. \ref{ssect:bar2}}, this
scenario cannot account for the large observed velocity splitting
along a single line of sight on the minor axis, at $\sim$ 3.5kpc from
the centre.

\citet{Athanassoula:2006} investigate the twist of isophotes of the
NIR 2MASS image, which shows a boxy shape bulge.  They compare this
morphology with simulated stellar models with bars, without gas.  They
conclude that the NIR morphology of M31 in the centre favours a bar,
seen almost side-on, i.e. the angle between the bar and the major axis
of the galaxy is between 20 and 30$^{\circ}$.  The size of the
boxy-bulge is 1.4\,kpc in radius, and the total length of the bar would
be 4\,kpc in radius, because a thin bar is identified beyond the boxy
bulge \citep{Beaton:2007}.

Let us note that in these two above propositions, the bar almost
end-on, or almost side-on, there is no large angle between the bar
and the symmetry axes of the projected galaxy (major or minor
axis). This is required because of the morphological symmetry of the
disc (no bar is obvious in the gas), and also because of the fairly
regular velocity field. If the bar were inclined by $\sim$ 45$^\circ$
with respect to the major axis, strongly skewed {(S-shape)} {
isovelocities} should be seen in the gas, i.e. the kinematical minor
axis would not be perpendicular to the major axis, as in N2683
\citep{Kuzio:2009}, which has been compared to M31. In the latter, owing
to the relative deficiency of gas in the centre
\citep{Brinks:1984b,Nieten:2006}, it is difficult to trace {
isovelocities}, but Figs. \ref{fig:XHaboul} and 
\ref{fig:velo-both} reveal that there is no strong skewness.

\citet{Athanassoula:2006} also {present} a 2D gaseous velocity field  in
an analytical bar potential (not fitted to M31) {and project it}
edge-on, {i.e. all regions of the galaxy up to 20\,kpc can thus be
seen artificially on the same line of sight, which makes a big
difference from a 77$^\circ$ projection corresponding to M31's real
configuration.}  The model 2D gas morphology clearly shows a bar, with
emptied regions that are not similar to the de-projected view of M31
gaseous disc \citep[e.g.][]{Braun:1991}.  The authors compare the
position-velocity diagrams parallel to the major axis that they
derived from the 2D-edge-on model to those observed in HI. The latter
reveal conspicuous figure-eight shapes that turn at the end of the HI
extent (R=70\arcmin $\sim$ 15\,kpc), and have been interpreted as
caused by the warp \citep{Henderson:1979,Brinks:1984a}. {These
features} are much larger than the {(3-4\,kpc)} bar extent. Note that
in the galaxy NGC 2683, as inclined on the sky plane as M31, the
position-velocity diagram along the major axis is not a figure-eight
shape, but a parallelogram that ends at a certain radius, precisely
the radius of the bar R= 2.2\,kpc= 45\arcsec
\citep{Kuzio:2009}. The situation is completely different in M31.

\subsubsection{Special case of M31}
\label{ssect:bar2}
Is it possible to obtain two velocity components on each side of the
systemic velocity along the same line of sight, near the minor axis
(position of M31-I and J), at a distance of 210\,arsec, i.e. 3.5\,kpc
from the centre (if the points are in the main inclined plane)?  The
possibility to gather several components in the same beam strongly
depends on the spatial resolution. Here, the CO(2-1) beam is 12\arcsec
$\sim$ 40\,pc on the major axis, however, the inclination of the plane
limits the resolution. Typically, gaseous discs in the centre of
galaxies have a characteristic scale-height of 50\,pc, as determined
in the Milky Way \citep[e.g.][]{Sanders:1984}. The inclined line of
sight across the disc of M31 encompasses a region of size
=tan(77$^\circ$) $\times$ 50\,pc= 216\,pc in the direction parallel to
the minor axis (while it is the size of the beam, 40\,pc, in the
perpendicular direction).  The region explored in one line of sight
then goes along the minor axis from 3.4\,kpc to 3.6\,kpc, and only
40\,pc in the perpendicular direction.  The only place where it would
be possible to have two velocity components on each side of the
systemic velocity is the very centre, because the 216\,pc region
explored { would encompass the two sides of an elongated orbit.}
Assuming that the bar is inclined with some angle with respect to the
major axis, the elongated orbits will provide red- and blue-shifted
velocities on each side of the centre. This is the case in NGC2683 for
instance, which is revealed by a long slit exactly along the major
axis. It is possible in the centre, because the bar is inclined with
respect to the symmetry axes. However, at a distance of 3.5\,kpc from
the centre ($\pm 100\,pc$), it is impossible to obtain these two
V-components, with the same category of orbits, either {\it x1} or
{\it x2}.  The only other possibility so far from the centre would be
to consider a change of orientation in the orbits, or a sudden shock,
as occurs from {\it x1} to {\it x2}, or crossing an arm. However, the
amplitude of the shock, 260\,km\,s$^{-1}$, and the positions of the
two components on the two sides of V$_{sys}$ is unrealistic,
especially for a possible weak bar.

\subsubsection{Examples of bar shocks}
\label{ssect:bar3}
Several strongly barred galaxies have been studied in H$\alpha$
spectroscopy, determining the kinematics with high spatial resolution,
comparable to what we have in the CO gas towards M31.  NGC\,1365 is
the prototype of a very strong bar. The position angle of the bar does
not coincide with the symmetry axis of the projection, so that the bar
signatures are obvious.  With an inclination of i=42$^\circ$, and a
distance of D=20\,Mpc, the observed velocity gradient across the bar
dust lanes is determined to be at maximum 40\kms, with a spatial
resolution below 100\,pc \citep{Lindblad:1996}.  If deprojected, the
maximum would be 60\kms.  NGC\,1300 is another remarkable and strongly
barred spiral galaxy, where the kinematics has been studied in detail
in H$\alpha$, on scales smaller than 100\,pc. \citet{Lindblad:1997}
present a projected velocity gradient across the dust lanes with
shocks of 20-30\,\kms projected on the sky.  When deprojected, this
becomes a maximum of 35-50\,\kms.  To reach these strong velocity
gradients, a highly contrasted gas structure must be found, which is
not the case for the M31 central region.

In the molecular component, interferometric observations of nearby
barred galaxies can now yield sub-arcsec resolution, corresponding to
scales smaller than 100\,pc, therefore comparable to the single-dish
resolution towards M31. \citet{Garcia-Burillo:2005} have presented
non-circular motions caused by bars, which are of the same orders of
magnitude $\sim$ 60-70\,\kms. More recently, with even higher
sensitivity and resolution, \citet{vanderLaan:2011} have for the first
time seen dedoubled velocity profiles in the CO data towards the
barred galaxy NGC 6951. If interpreted in terms of crowding of orbits,
they reveal velocity differences of 80\,\kms {(within 500\,pc)},
which in deprojection would amount to 110\,\kms. The CO velocity
profiles are, however, very broad in the centre of this galaxy,
showing a starburst ring, and it is not sure what is caused by turbulence
or by the bar. The molecular gas surface density in the centre is on
average 100\,M$_\odot$/pc$^2$ {(corresponding to a total mass of
$1.5 \times 10^9\msol$ inside 2\,kpc)}, which may explain the unstable and
turbulent gas-forming stars, while in M31 the surface density is at
least one order of magnitude lower.  The velocity dispersion in the
distinct velocity components observed in CO is quite low (15\,\kms),
which corresponds to a stable gas layer. We conclude that the bar is
not a likely explanation for the special kinematical phenomenon
observed towards the M31 centre.

\subsubsection{Other interpretations}
\label{ssect:bar4}
 From the HI-derived gas morphology,
\citet{Braun:1991} suggests a two-armed trailing density wave with a
pattern speed of 15\,\kms\,kpc$^{-1}$, leading to a corotation
at 16\,kpc, an ILR at 5\,kpc, and an OLR at 22\,kpc. He proposes as the
driver of the spiral structure the collision with M32, which could
also be responsible for the many tilts of the M31 plane.  The nearly
head-on collision with M32 has been developed in detail by
\citet{Block:2006}, which allows them to account for the two
off-centred gas rings observed in M31 at 0.7\,kpc and 10\,kpc
radius. Indeed, these authors note that the gas morphology in M31 does
not show the usual signatures of the response to a bar. Although it is
possible that the galaxy disc possesses a stellar bar, aligned with
the triaxial bulge, oriented roughly along the major axis, the
observed 0.7\,kpc dust ring is aligned almost along the minor axis,
and not aligned with the potential bar along the major axis.  The size
of the inner ring does not correspond to a possible inner Lindblad
resonance, nor the outer ring at 10\,kpc radius to a correlated outer
resonance. Moreover, the inner disc appears highly inclined with
respect to the main disc, and the inner ring is off-centred by 40$\%$
of its radius, which is not expected in barred galaxies. This offset
centre and the tilted plane of the inner disc strongly point towards a
perturbed origin of the inner features of M31.

\subsection{Scenario 3: head-on collision with M32}
\label{ssect:ring}
\citet{Jacoby:1985} {have shown} that the inner disc of M31 is
tilted with respect to the main disc. While the inclination of the
large-scale disc is 77$^\circ$, the nuclear disc, of size $\sim$
1\,kpc, is inclined by only about 40$^\circ$ on the sky plane.
\citet{Block:2006} discovered in the {\em Spitzer}-IRAC maps an
inner dust disc of scale 1\,kpc by 1.5\,kpc, which also appears with
this low inclination. The ring morphology of this tilted structure
together with the ring (10kpc) morphology of the large-scale disc led
these authors to propose a head-on collision with a small companion, with a
probable candidate being M32.

In this scenario, we assume that the peculiar component comes from a
tilted ring-like material, likely coming from the perturbed gas from
the recent M32 collision.  The collision might have perturbed the
nuclear disc, which was already tilted, and produced some local warp
before the annular density wave propagated outwards.  {Besides its
interest for the inner configuration of M31, this scenario also
proposes an explanation for the 10\,kpc ring, resulting from the
propagation of the initial annular wave.}

The resulting configuration is similar to the scenario proposed in
Sect. {\ref{ssect:loop}}. Below, we discuss some modelling that
enables one to explain the observations. We stress that this modelling
is meant to be schematic rather than reproducing the observations in a
fully self-consistent manner.

We used a model for the gas, through static simulations of dynamical
components represented by particles, with different geometrical
orientations on the sky, but embedded in a gravitational potential
representing the observed rotation curve of M31.

We represent the gaseous disc of M31 by a fairly homogeneous
Miyamoto-Nagai disc of particles (with radial scale of 1\,kpc, and
height of 0.2\,kpc) to be able to vary easily the inclination and
position angle as a function of radius.  We used typically half a
million particles to have sufficient statistics.  We plunged the gas disc
into a potential made of a stars and a dark matter halo.  The stellar
component is composed of a bulge and a disc.  The bulge is initially
distributed as a Plummer sphere, with a potential
\begin{equation}
\Phi_{b}(r) = - { {G M_{b}}\over {\sqrt{r^2 +r_{b}^2}} },
\end{equation}
where  $M_{b}$ and  $r_{b}$ are the mass and characteristic radius of the
bulge, respectively (see Table \ref{condini}).

The stellar disc is initially a Kuzmin-Toomre disc of surface density
\begin{equation}
\Sigma(r) = \Sigma_0 ( 1 +r^2/r_d^2 )^{-3/2}
\end{equation}
with a mass  $M_d$,
and characteristic radius $r_d$.

The dark matter halo is also a Plummer sphere, with mass $M_{\rm DM}$
and characteristic radius $r_{\rm DM}$.  A summary of the adopted
parameters is given in Table \ref{condini}.  The resulting rotational
velocity curve is rising until a maximum of 300\kms, reached at
3\,kpc, and then slightly decreases, remaining close to 300\kms\, in
the central regions of interest here. This corresponds to the
rotational velocity observed
\citep[e.g.][]{Carignan:2006}. All molecular clouds we aim to
reproduce are inside the radius of maximum rotational velocity.

The gas particles are distributed in this potential, with the velocity
dispersion corresponding to a Toomre $Q$ parameter of $1.3$.

\begin{table}[ht]
\caption[ ]{Model parameters}
\begin{flushleft}
\begin{tabular}{ccccccc}  \hline
Component       & radius & Mass & Mass fraction   \\
                          & [kpc]  & [$M_\odot$]&    [ \%]                \\
\hline
Bulge          &  0.2    &  1.6e10      &  4.5       \\
Disc             & 4.0   &  7.e10      &   20.       \\
Halo        &  10.   & 27.e10    &   75.5     \\
\hline
\end{tabular}
\end{flushleft}
\label{condini}
\end{table}
\begin{figure}[ht]
\centering
\includegraphics[angle=-90,width=8.5cm]{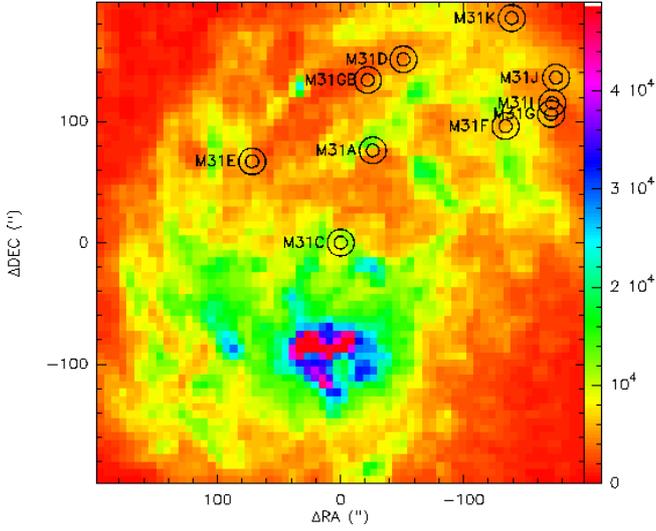}
\caption{ Mean density of CO emission from the homogeneous nuclear disc plus tilted ring of
the model. The field of view is 200 arcsec in radius, or 0.76\,kpc in radius. The density scale, indicated in the wedge, is in arbitrary units.
{The positions of our CO observations are indicated with circles corresponding to the beams.}}
\label{mean}
\end{figure}
\begin{figure}[ht]
\centering
\includegraphics[angle=-90,width=8.5cm]{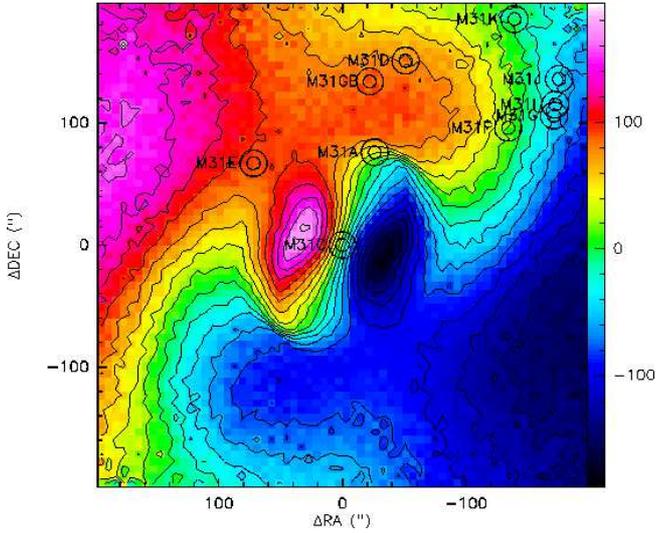}
\caption{{ Density-weighted mean velocity (first moment)} of the
simulated nuclear region  in the same field of view 
as Fig \ref{mean}.  Signatures of the tilted ring can be seen at the
boarder of the field. The wedge gives the velocity scale in \kms.}
\label{velo}
\end{figure}
\begin{figure}[ht]
\centering
\includegraphics[angle=-90,width=8.5cm]{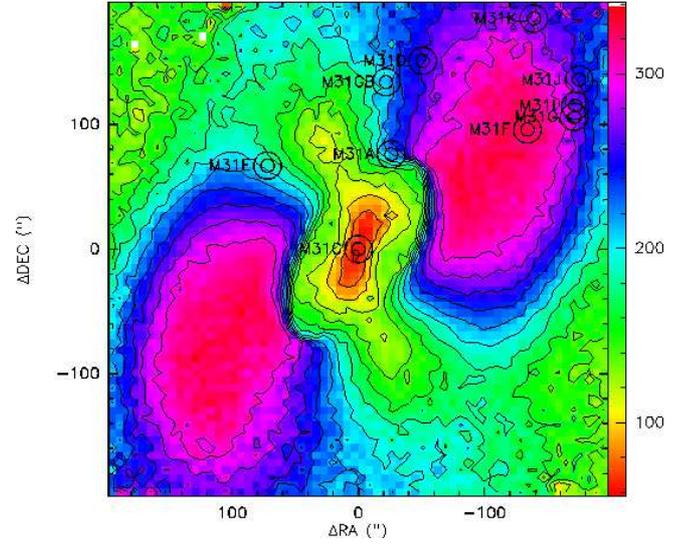}
\caption{{ Density-weighted velocity dispersion (second moment of
the simulated velocity field).}  The wedge gives the velocity scale in
\kms. The locations of double components with counter-rotating
features correspond to the blue and purple regions. }
\label{width}
\end{figure}
To reproduce the observations, we adopted an inclination of 43$^\circ$
for the gaseous nuclear disc, inside the 1.5\,kpc radius. At the
boundary of this disc, we assumed a progressive warping of the disc
plane, so that the inclination on the sky grows from 43 to
77\,degrees, over 300\,pc.  The details of this transition, however,
are not constrained by the observations, because the latter are all
inside the 1\,kpc radius.  As for the position angle on the sky, the
nuclear disc has PA$=70^\circ$, unlike the main disc, which has
PA$=35^\circ$.  In projection on the sky, the nuclear gas disc model
gives velocities that agree with the observations, at least with the
main velocity component.  In many observed points only this main
component is observed, with a broad line-width (FWHM$=$50-70\,\kms),
located on the blueshifted side in the SW and on the redshifted side
in the NE of the major axis.  In some of the points there is an
additional peculiar velocity component, located on the opposite side
(redshifted in the SW), and narrower (20 \kms).

Trying several parameters, we found that material on a ring-like
orbit, with 40$^\circ$ inclination, and PA= -35$^\circ$ gives
kinematical features compatible with the observations (cf Figure
\ref{mean}-\ref{spec}). The modelled ring has a width of 0.4\,kpc, and
a mean radius of 0.6\,kpc, coinciding with the observed dust ring.  The
adopted position angle and inclination of the various components are
displayed in Table \ref{geometry}.
\begin{table}[ht]
\caption[ ]{Geometrical parameters}
\begin{flushleft}
\begin{tabular}{ccccccc}  \hline
Component       & Main disc & Nuclear disc & Inner ring   \\
                          & [$^\circ$]  & [$^\circ$]&    [ $^\circ$]                \\
\hline
\hspace{0.2cm}
PA                     &  35   &   70 &  -35       \\
Inclination            & 77    &  43       &   40      \\
\hline
\end{tabular}
\end{flushleft}
\label{geometry}
\end{table}
\begin{figure}[ht]
\centering
\includegraphics[width=8cm]{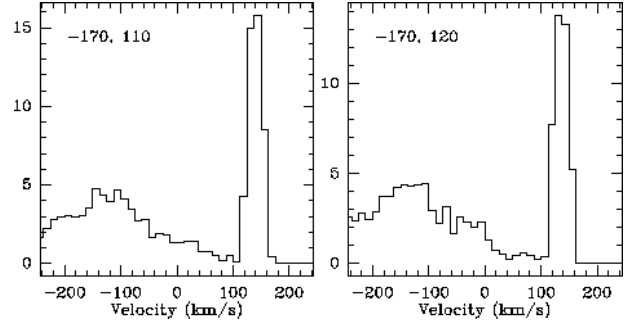}
\caption{Spectra extracted from the simulated cube at the (RA, DEC)
offset in arcseconds indicated in each panel. The vertical scale is
in arbitrary units. The velocities are centred at the systemic
velocity. These offsets correspond the M31G and M31I positions. 
The clumpiness of the gas (not included in the modelling proposed
here) is expected to  modulate the relative intensity of the
various components additionally.}
\label{spec}
\end{figure}

To draw these maps, we built data cubes corresponding to the
observations, with a pixel of 6 arcsecond, and channels of 12 \kms,
and the data smoothed to a beam of 12\arcsec.  The {pixel} size of the
cubes are therefore (60, 60, 40) to describe the inner parts studied
here in CO lines and dust extinction. To take into account that the
gas and dust distribution is not homogeneous, but patchy, we smoothed
the dust map obtained by \citet{Block:2006} from {\em Spitzer}-IRAC
images (the 8\mic\, map with the 3.6\mic\, contribution subtracted) to
the same spatial resolution, and used it as a multiplicative filter
for our homogeneous disc map.  This takes into account that the dust
inner ring is offset by 0.5\,kpc (131'') from the centre of the
galaxy.

Typical spectra were extracted in the zone where double features are
expected at the offset indicated in Fig \ref{spec} in arcseconds.  On
the one hand, the broad characteristics of the blue-shifted velocity
component is caused by beam smearing of the velocity gradient, as
illustrated here with a 12$\arcsec$ beam.  On the other hand, because
the ring is narrow in radius ($\sim$ 170\,pc) and in a region of a
flat rotation curve, the velocity gradient in the beam is very flat
and a narrow line is observed in the redshifted component.

In addition, some emission at the systemic velocity is expected
because of the inclination of the 77$\deg$ main disc. It is therefore
not necessary to evoke the large-scale warp to explain the weak
emission at the systemic velocity, which can be accounted for by the
inner main disc. For positions like M31I/M31J/M31G located in the
north-west ring, main disc regions at 3.5\,kpc project in the same
beam.

\section{Discussion and conclusions}
\label{sect:disc}
\begin{figure} \centering
\includegraphics[height=7cm]{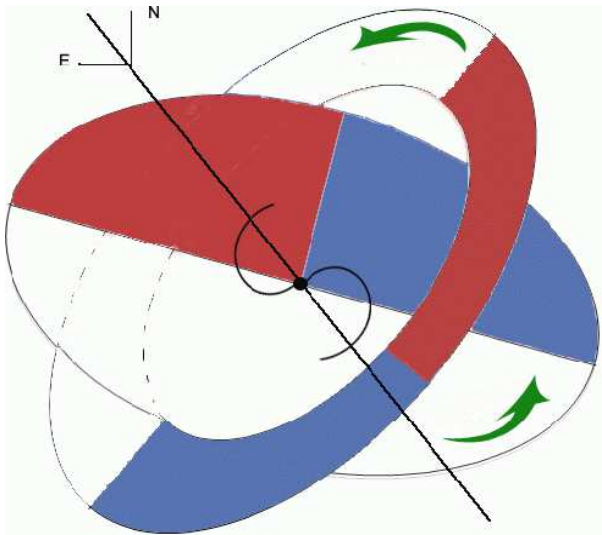}
\caption{Schematic view of the interpretation proposed for the CO velocities
observed. The inner disc is presented with a PA of 70$\deg$ and an
inclination of 43$\deg$. The inner ring is superimposed with a similar
inclination but a position angle of -35$\deg$. The straight line
indicates the position of the major axis of the main disc inclined by
77$\deg$ with a PA of 35$\deg$.  The bars labelled ''N'' and ''E''
have a length of 150\,pc. { The near (resp. far) sides of the two
components are coloured (resp. kept white). The red (resp. blue)
colours correspond to the redshift (resp. blueshift) relative to the
systemic velocity.}}
\label{fig:schem}
\end{figure}
We have detected CO emission, tracing the presence of molecular gas in
several positions in the central kpc of M31, corresponding to dust
extinction. {We have also shown that there is an excellent
correspondence between the dust extinction measured in the optical and
the dust emission in the near-infrared at 8$\mu$m (right panel of
Figure \ref{fig:XHa}).} Some of the {observed} positions correspond to
the conspicuous inner ring, detected in dust emission by IRAC on {\em
Spitzer} \citep{Block:2006}}. This ring was also present in ISO data
studied by \citet{Willaime:2001} and in ionised gas maps as proposed
by \citep{Jacoby:1985}. The velocity information brought by the CO
lines reveals a complex kinematical structure, different from the
ionised {and atomic gas dynamics}. In the NW inner ring positions, two
{main high S/N} velocity components are detected on each side of the
systemic velocity.  The component with the expected velocity,
according to the rotation curve and the positive position angle, is
broad, and is likely to correspond to the nuclear disc. Exploring a
wide region of the strong central rotational gradient, the
beam-averaged spectrum may be 100km/s broad. The second component,
redshifted on the other side of the systemic velocity, has a small
velocity width, because it is not a full disc, but a relatively narrow
ring.  We used radii from 0.4 to 0.8\,kpc for the model.

We summarise in Figure \ref{fig:schem} the geometry { of the two
components} we propose in our modelling.  The peculiar component
appears to be counter-rotating, because the gas is in a tilted ring,
almost perpendicular to that of the inner disc. Its inclination on the
plane of the sky is similar, which explains why the amplitude of the
projected rotation is the same as for the regular component. In the
figure the arrows indicate the apparent sense of rotation. The two
discs appear to rotate in the same direction, which is also that of
the main M31 disc.  The apparent counter-rotation appears only in the
NW (and SE) regions, where blue and red regions superpose.  The
winding sense of the spiral structure in the observations is also
sketched. It is possible to see that the arms are trailing, exactly
like the M31 main disc.

In summary, the apparent counter-rotation is only caused by a warping
and distortion of the central components, possibly triggered the
proposed head-on collision with M32, but is not a true
counter-rotation.

\begin{acknowledgements}
We are most grateful to F. Viallefond, who designed the observing
procedures and to M. Marcelin, who retrieved for us old kinematical
$[$N{\sc ii}$]$ data. We thank R. Saglia who has kindly provided the
slit spectra he took in the bulge, enabling us to search for double
components and {\'A} Bogd{\'a}n for providing their reduced XMM-Newton
and Chandra maps of the bulge exhibiting an X-ray outflow and helpful
comments.  We acknowledge I. Chilingarian for his help in the spectra
reduction and M. Sarzi for helpful comments.  {We are very grateful to
Robert Braun for his constructive comments, and for having provided
the HI data, in order to check the HI correlation with the CO data.}
We thank the anonymous referee for his constructive remarks, which
helped us to improve the manuscript.  This paper is based on
observations carried out with the IRAM 30m telescope. IRAM is
supported by INSU/CNRS (France), MPG (Germany) and IGN (Spain).  The
authors are grateful to the IRAM staff for their support.
%and to the referee for helpful comments.
\end{acknowledgements}

\newpage
\appendix
\section{CO signal detected in the OFF positions}
\label{sect:COOFF}
\begin{figure}
\centering 
   \includegraphics[width=6.5cm,angle=-90]{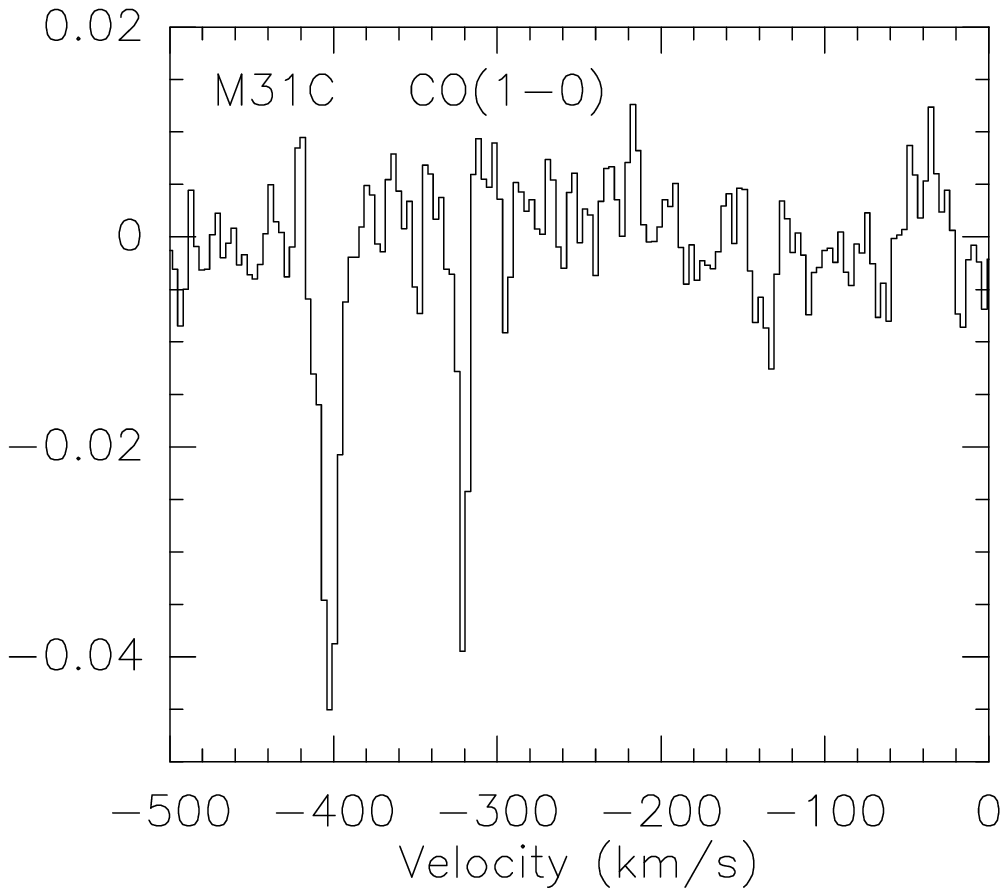} \caption{
   Signal detected in 2000 at the J2000 coordinates RA 00:43:18.5 DEC
   +41:17:06, in the OFF signal of position switch observations of
   M31C. This spectrum, corresponding to an exposure time of 80\,min,
   was reduced to a 24$^{\prime\prime}$ beam. The y axis displays
   antenna temperature T$_A^*$.}  \label{fig:AppFig1}
\end{figure}
\begin{figure}
\centering 
   \includegraphics[width=6.5cm,angle=-90]{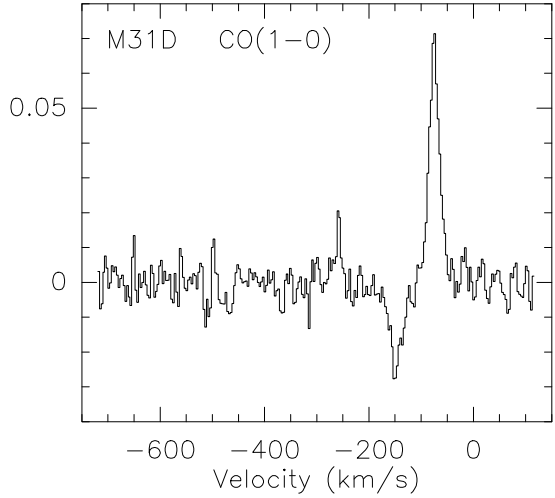} \caption{
   Signal detected in the OFF positions of M31D observations, with the
   wobbler switch observing mode, in 1999 (wobbler throw:
   157$^{\prime\prime}$- 165$^{\prime\prime}$).  The y axis displays
   antenna temperature T$_A^*$.} \label{fig:AppFig2}
\end{figure}
\begin{figure}
\centering 
   \includegraphics[width=6.5cm,angle=-90]{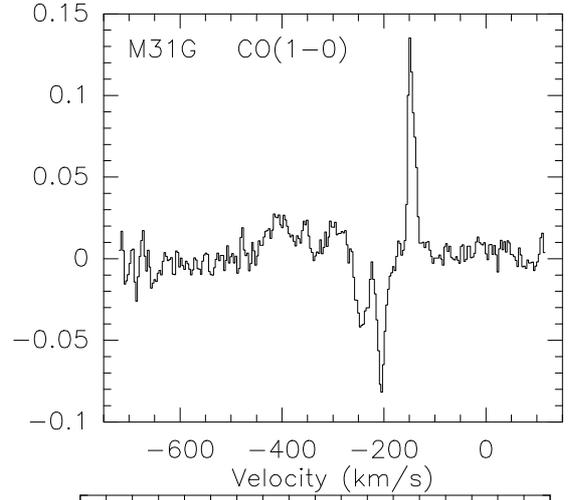}
   \includegraphics[width=6.5cm,angle=-90]{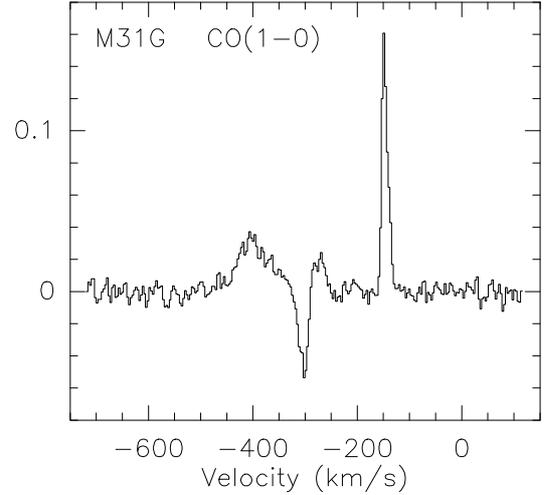} \caption{
   Signal detected in the OFF positions of M31G observations, with the
   wobbler switch observing mode, in 1999 (top: wobbler throw
   75$^{\prime\prime}$- 90$^{\prime\prime}$) and in 2000 (bottom:
   wobbler throw 94$^{\prime\prime}$ - 104$^{\prime\prime}$). The
   2000 observations (bottom panel) was reduced to a 24$^{\prime\prime}$
   beam.  The y axis displays antenna temperature T$_A^*$.}
   \label{fig:AppFig34}
\end{figure}
 We provide here a short description of the CO signal detected in the
 OFF positions of the IRAM observations presented in this paper. Their
 interpretation is beyond the scope of this paper, because they
 explore a larger area than the ON positions and are only partial.
\subsection*{OFF position of M31C (position switch)}
 A clear signal was detected at $V_0=-402.3$\,km\,s$^{-1}$
($\sigma=12.5$\,km\,s$^{-1}$, $I_{CO}=0.85$\,K\,km\,s$^{-1}$) and
$V_0=-320$\,km\,s$^{-1}$ ($\sigma=5.7$\,km\,s$^{-1}$,
$I_{CO}=0.36$\,K\,km\,s$^{-1}$) in the OFF position of the position
switch observations of M31C, as displayed in Figure
\ref{fig:AppFig1}.

\subsection*{OFF position of M31D (wobbler switch)}
For M31D no OFF signal was detected in the position switch scans
performed at the OFF position
122$^{\prime\prime}$,321$^{\prime\prime}$. However, there is a
detectable OFF signal for the scans obtained in wobbler switch mode in
1999, as displayed in Figure \ref{fig:AppFig2}. This signal
corresponds to a wobbler throw centred with 157$^{\prime\prime}$ and
165$^{\prime\prime}$ from the ON position. Because the observing procedure
was designed to avoid extincted areas and the area surrounding M31D
was not crowded, we suspect that this OFF signal close in velocity to
the ON signal is caused by gas on the back side (not detected in
extinction). However, complementary observations are required to
confirm this hypothesis. The detected signal is at
$V_0=-147.7$\,km\,s$^{-1}$ ($\sigma=31.42$\,km\,s$^{-1}$,
$I_{CO}=1.12$\,K\,km\,s$^{-1}$) with an exposure time of 162\,min.

\subsection*{OFF position of M31G (wobbler switch)}
No OFF signal was detected in the position switch scans observed for
M31G in 2000 and only wobbler mode has been used in 1999. However, a
signal was observed in the OFF positions obtained with the wobbler
switch mode, as displayed in Figure \ref{fig:AppFig34}. The throw
positions were different in 1999 and 2000 and the OFF signal detected
was also different.  In 1999 the detected signals are at
$V_0=-204.5$\,km\,s$^{-1}$ ($\sigma=17.5$\,km\,s$^{-1}$,
$I_{CO}=2.07$\,K\,km\,s$^{-1}$) and $V_0=-242.6$\,km\,s$^{-1}$
($\sigma=23.2$\,km\,s$^{-1}$, $I_{CO}=1.42$\,K\,km\,s$^{-1}$)
corresponding to an exposure time of 180\,min. In 2000 the OFF signal
is detected at $V_0=-303.3$\,km\,s$^{-1}$
($\sigma=17.6$\,km\,s$^{-1}$, $I_{CO}=1.36$\,K\,km\,s$^{-1}$),
corresponding to an exposure time of 142\,min.  As discussed in the
paper, these OFF positions superimpose on a wide ON signal. Because we do
not know the position of the wobble throw (only its distance from the
ON position), we can only argue that these OFF signals detected within
1.8\,arcmin from the ON position can be associated to the same
structure.

\section{HI spectra}
\label{sect:append}
\begin{figure*} \centering
\includegraphics[height=15cm,angle=-90]{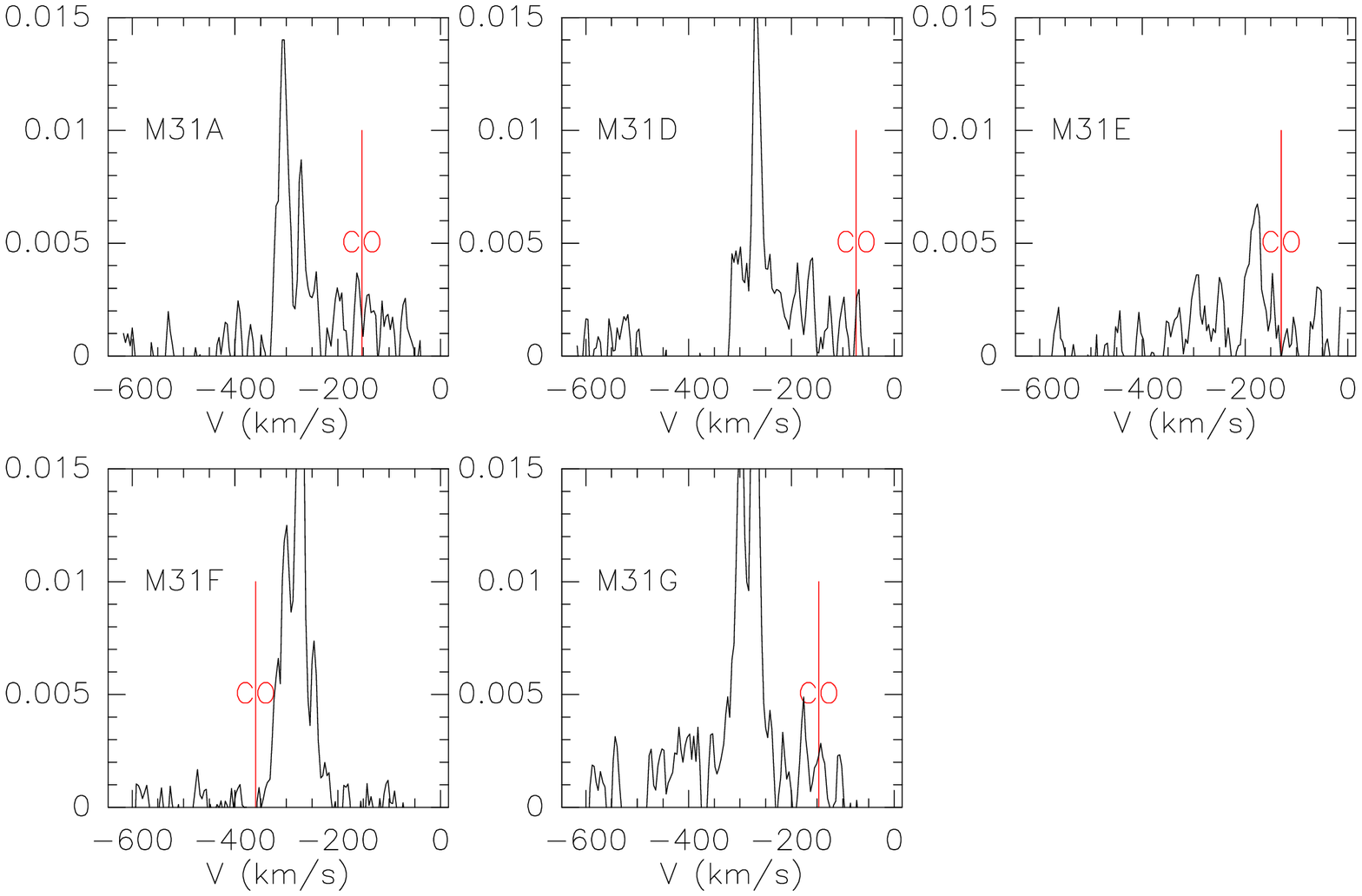}
\includegraphics[height=15cm,angle=-90]{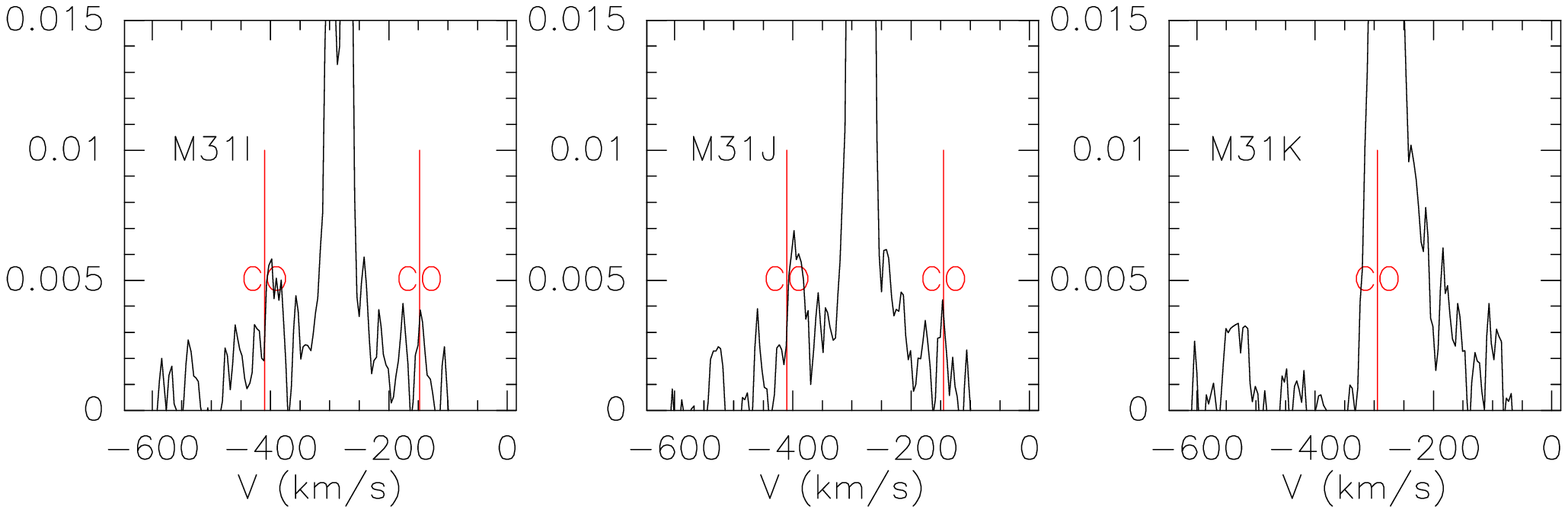}
\caption{HI spectra obtained with a 1 arcmin resolution for the CO
positions discussed in this paper. In each spectrum the velocity of the
CO emission is indicated in red. }  
\label{fig:HIvel}
\end{figure*}
Figure \ref{fig:HIvel} displays the HI velocities obtained for the
different positions observed in CO. The positions M31I and M31J
exhibit one possible component compatible with the disc component
detected in CO. There is no clear HI counter-part for the ring component.

\end{document}